%% file: main.tex
\documentclass[sigconf, nonacm]{acmart}
\usepackage[font=small,skip=0pt]{caption}
\usepackage{graphicx}
\usepackage{textcomp}
\usepackage{xcolor}
\usepackage{listings}
\usepackage{algorithm}
\usepackage[noend]{algpseudocode}
\usepackage{tikz}

\usepackage{multirow}
\DeclareMathOperator*{\minimize}{minimize}

\usepackage{amsmath}

\usepackage{hyperref}
\usepackage{enumitem,kantlipsum}

\hypersetup{
  colorlinks,
  citecolor=violet,
  linkcolor=red,
  urlcolor=blue}

\newcommand*\circled[1]{\tikz[baseline=(char.base)]{
            \node[shape=circle,fill,inner sep=2pt] (char) {\textcolor{white}{#1}};}}

\newtheorem{thm}{Theorem}[section]

\theoremstyle{definition}
\newtheorem{definition}{Definition}[section]

\newcommand{\showComments}{yes}
\newcommand{\submit}{no}            

\newcommand{\note}[2]{
  \ifthenelse{\equal{\submit}{yes}}{}{%
    \ifthenelse{\equal{\showComments}{yes}}{\textcolor{#1}{#2}}{}
  }
}

\usepackage{dsfont}
\newcommand{\todo}[1]{\note{red}{[TODO: #1]}}
\newcommand{\xiao}[1]{\note{green}{[Xiao: #1]}}

\newcommand{\review}[1]{\note{cyan}{[Review: #1]}}

\AtBeginDocument{%
  }

\setcopyright{acmlicensed}
\copyrightyear{2018}
\acmYear{2018}
\acmDOI{XXXXXXX.XXXXXXX}

\acmConference[Conference acronym 'XX]{Make sure to enter the correct
  conference title from your rights confirmation emai}{June 03--05,
  2018}{Woodstock, NY}
\acmISBN{978-1-4503-XXXX-X/18/06}

\begin{document}

\title{Privacy and Accuracy-Aware AI/ML Model Deduplication
}
\subtitle{Accepted by SIGMOD 2025. This is a pre-print version.}

\newcommand{\tsc}[1]{\textsuperscript{#1}}

\author{Hong Guan\tsc{1}, Lei Yu\tsc{2}, Lixi Zhou\tsc{1}, Li Xiong\tsc{3},\\Kanchan Chowdhury\tsc{1}, Lulu Xie\tsc{1}, Xusheng Xiao\tsc{1}, Jia Zou\tsc{1}}
\affiliation{
  \institution{\tsc{1} Arizona State University, \tsc{2} Rensselaer Polytechnic Institute, \tsc{3} Emory University}
  \country{}
}
\renewcommand{\shortauthors}{Guan et al.}
\begin{abstract}
%
With the growing adoption of privacy-preserving machine learning algorithms, such as Differentially Private Stochastic Gradient Descent (DP-SGD), training or fine-tuning models on private datasets has become increasingly prevalent. This shift has led to the need for models offering varying privacy guarantees and utility levels to satisfy diverse user requirements. However, managing numerous versions of large models introduces significant operational challenges, including increased inference latency, higher resource consumption, and elevated costs.
Model deduplication is a technique widely used by many model serving and database systems to support high-performance and low-cost inference queries and model diagnosis queries. However, none of the existing model deduplication works has considered privacy, leading to unbounded aggregation of privacy costs for certain deduplicated models and inefficiencies when applied to deduplicate DP-trained models. 

We formalize the problems of deduplicating DP-trained models for the first time and propose a novel privacy- and accuracy-aware deduplication mechanism to address the problems. We developed a greedy strategy to select and assign base models to target models to minimize storage and privacy costs. When deduplicating a target model, we dynamically schedule accuracy validations and apply the Sparse Vector Technique to reduce the privacy costs associated with private validation data. 
Compared to baselines that do not provide privacy guarantees, our approach improved the compression ratio by up to $35\times$ for individual models (including large language models and vision transformers). We also observed up to $43\times$ inference speedup due to the reduction of I/O operations. 
%
\end{abstract}

\maketitle

\input{intro}

\input{background}

\input{problem}
\input{overview}

\input{implementation}

\input{experiments}
\input{discussion}
\input{relatedworks}
\input{conclusion}

\bibliographystyle{ACM-Reference-Format}
\bibliography{refs}

\end{document}

%% file: intro.tex
\vspace{-10pt}
\section{Introduction}
\textcolor{black}{
AI/ML systems that support model marketplace~\cite{liu2021dealer, lin2014arbitrage, chen2019towards}, ML-as-a-Service (MLaaS)~\cite{luo2021privacy, hunt2018chiron, yin2024llm}, and multi-tasking models on edge devices~\cite{lee2020fast}, require efficient management of many models, including large language models (LLMs). These models are often trained or fine-tuned on private datasets using privacy-preserving algorithms~\cite{yu2021large, yu2021not}, such as DP-SGD~\cite{abadi2016deep}. Such algorithms are widely used to  protect training data from attacks, such as re-identification~\cite{benitez2010evaluating}, membership inference~\cite{hu2022membership, shokri2017membership}, and model inversion~\cite{fredrikson2015model}. 
The management of many DP models poses new challenges:}

\textcolor{black}{
\noindent
$\bullet$ Model marketplaces~\cite{liu2021dealer, lin2014arbitrage, chen2019towards} train many versions of a model using different privacy budgets ($\epsilon$). Then, these models are sold to different buyers so that models with higher $\epsilon$ (less noise) and higher accuracy will require a higher price, providing more options to potential buyers, while ensuring arbitrage freeness that no model buyers can take advantage of the marketplace by combining multiple lower-quality models sold at lower prices into a higher-quality model that has a designated higher price~\cite{liu2021dealer, lin2014arbitrage}. In addition, each buyer or buyer group should be assigned a maximal accumulative $\epsilon$ over a private dataset based on what they pay or their privilege. This is widely adopted in many systems such as Google BigQuery~\cite{google-analysis-rules}, Snowflake~\cite{snowflake-privacy},  and DProvDB~\cite{zhang2023dprovdb}. Usually, a model needs to be stored after being sold for customer service and auditing, incurring operational costs.}

\textcolor{black}{
\noindent
$\bullet$ MLaaS platforms~\cite{luo2021privacy, hunt2018chiron, yin2024llm} have similar problems, serving many models trained with various $\epsilon$ values~\cite{xu2017advances,desfontaines2021list,mcsherry2009differentially}, while each customer's total privacy cost is limited by her/his privacy budget.  In addition, model serving is usually subject to stringent latency requirements defined in service-level objectives (SLOs)~\cite{crankshaw2019design}. 
When the models to be served exceed available memory, heavy I/O operations are required to swap models between disk and memory, making it even more challenging to meet the latency SLOs.}

\textcolor{black}{
\noindent
$\bullet$ Multi-tasking applications on edge devices~\cite{lee2020fast} deploy multiple DP-protected models (purchased from model marketplaces) in a resource-constrained environment, facing difficulties in balancing response latency and resource (memory/battery) limitations.  
}

 Model deduplication is promising to resolve these problems by sharing similar weight blocks across multiple models to dramatically reduce memory footprint and I/O operations, and thus shorten the latency for serving multiple models in resource-constrained environments~\cite{lee2020fast}, multi-tenant environments~\cite{crankshaw2017clipper}, and database systems that support inference queries~\cite{zhou2022serving} and model diagnosis queries~\cite{vartak2018mistique}. \textcolor{black}{In addition, as observed by prior works~\cite{lee2020fast} and illustrated in Fig.~\ref{fig:disparity}, the block-wise disparity score~\cite{lee2020fast} is small for models trained with different $\epsilon$ using DP-SGD, which also suggests that model deduplication is promising for addressing the target problems.} Unlike \textit{single-model} compression techniques such as quantization~\cite{gholami2022survey}, pruning~\cite{blalock2020state}, and knowledge distillation~\cite{gou2021knowledge}, deduplication is a \textit{multi-model} technique and can work together with single-model compression techniques by deduplicating multiple compressed models or compressing deduplicated models~\cite{zhou2024serving}.

{\color{black}
 \begin{figure}[ht]
\centering
\vspace{-3pt}
\includegraphics[width=0.4\textwidth]{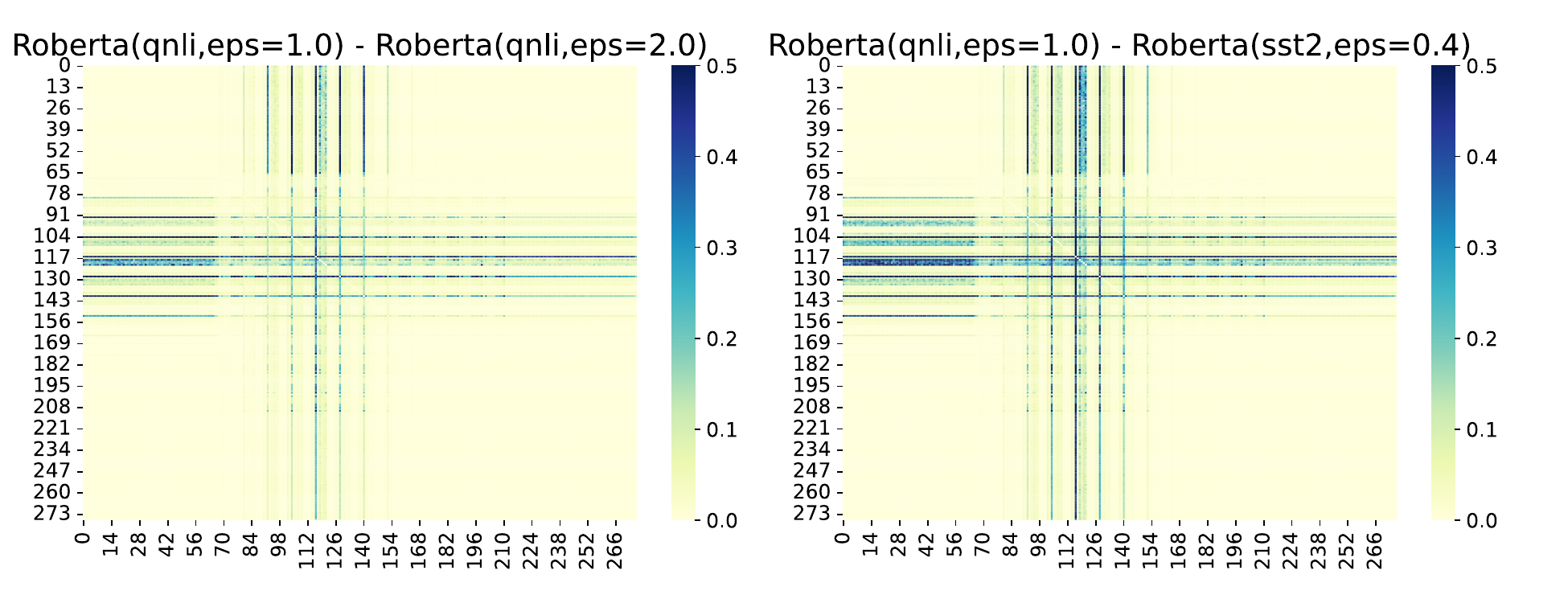}
\caption{\textcolor{black}{Weight disparity between DP-SGD-trained models. Left: RoBERTa finetuned on QNLI, with $\epsilon=1.0$ vs. $\epsilon=2.0$; right: RoBERTa finetuned on QNLI with $\epsilon=1.0$ vs fine-tuned on SST2 with $\epsilon=0.4$). The x- and y-axis represent blocks of two models respectively. Each point represents the disparity score~\cite{lee2020fast} between these blocks. 
}} 
\label{fig:disparity}
\vspace{-10pt}
\end{figure}
}

 \begin{figure*}[h]
\centering
\vspace{-5pt}
\includegraphics[width=1\textwidth]{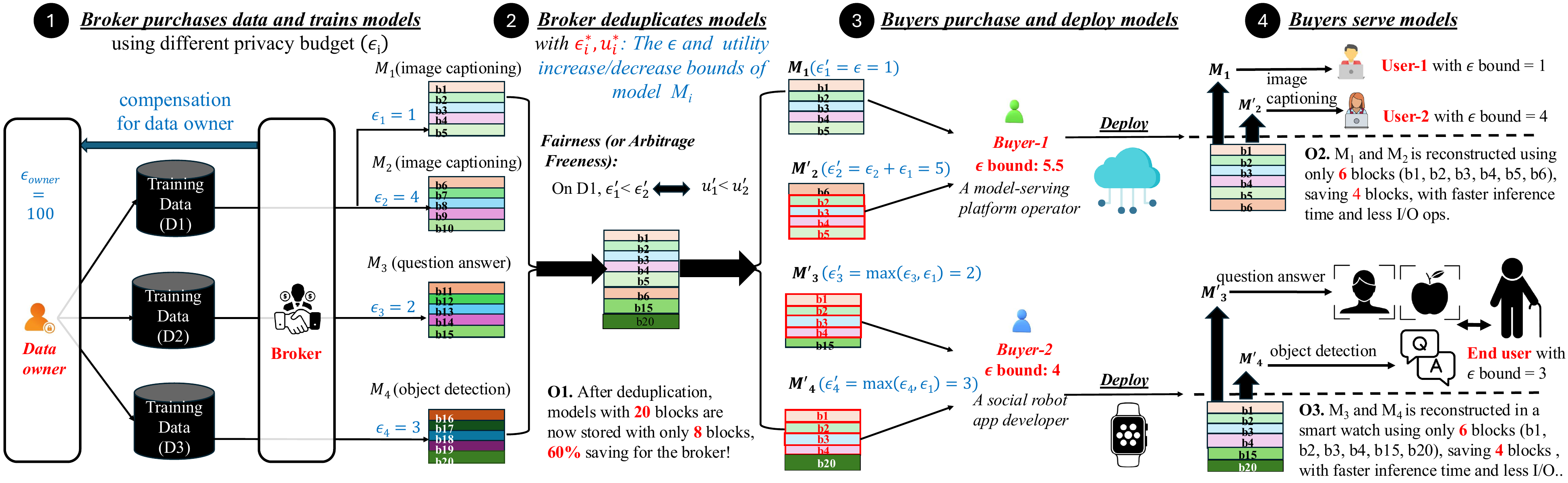}
\caption{ \textcolor{black}{Model Deduplication Example. $M_1$ (base model) trained on D1 provides blocks for replacing the similar blocks in $M_i$ with privacy and utility bound (e.g., $\epsilon'_i<\epsilon_i+\epsilon^*_i$). $\epsilon'_2$ follows sequential composition since $M_2$ is also trained on D1, while $\epsilon'_3$ and $\epsilon'_4$ follow parallel composition (See Sec.~\ref{sec:preliminary}). Each dataset (D1, D2, or D3) is a disjoint partition of a logical data collection, for which privacy loss should be assessed collectively.}}
\label{fig:deduplication-overview}
 \vspace{-10pt}
\end{figure*}

\textcolor{black}{In a block-based model deduplication scenario, as illustrated in Fig.~\ref{fig:deduplication-overview}, the models' weights are partitioned into uniform-sized blocks, e.g., $20$ blocks in Fig.~\ref{fig:deduplication-overview}\circled{1}. Blocks from a base model (i.e., $M_1$) are identified to replace blocks in other models (i.e., $M_2$, $M_3$, and $M_4$). As a result, only $8$ distinct blocks are needed to reconstruct the models (without significantly degrading models' utilities), which alleviated the broker's operational cost challenges (O1 in Fig.~\ref{fig:deduplication-overview}\circled{2}). 
However, none of the existing model deduplication works considered the privacy of the models, which poses significant new issues. 
%
If any of a model's blocks are replaced (i.e., deduplicated) by blocks from other models trained on the same dataset, the overall privacy budget incurred on the training dataset remains unchanged due to the post-processing rule (Sec.~\ref{sec:preliminary}), which means the total privacy loss on the training datasets will not change, and thus model deduplication will not affect data owners. However, the privacy loss ($\epsilon$) of a deduplicated model may increase (e.g., $\epsilon_2$ increases to $\epsilon'_2$ in Fig.~\ref{fig:deduplication-overview}\circled{2}) due to the composition of DP (Sec.~\ref{sec:preliminary}), which should be bounded based on its target user's privilege and trust-level on the dataset~\cite{zhang2023dprovdb, liu2021dealer}, for the following reasons:
(1) The private information leakage to a specific user may increase, which should be bounded by the user's authorized overall $\epsilon$.  For example, the increased privacy cost of $M_2$ ($\epsilon'_2$) causes more information leakage to Buyer-1 in Fig.~\ref{fig:deduplication-overview}\circled{3}, while Buyer-1 has an overall $\epsilon$ limit of $5.5$. It will further affect User-2 in Fig.~\ref{fig:deduplication-overview}\circled{4}, the end user of $M_2$, with an $\epsilon$ limit of $4$.
 (2) The deduplication may affect both the model $\epsilon$ and accuracy, which should satisfy \textit{the (arbitrage-free) fairness rule} that a model trained on the same dataset having a higher accuracy must have a higher $\epsilon$ ~\cite{liu2021dealer}. We further consider that the $\epsilon$ increase of a model caused by deduplication must be \textbf{minimized} so that its buyer can save some privacy budget for purchasing more models.}

In light of this scenario, we define a \textbf{novel deduplication problem}: to optimize the storage and privacy costs of all DP-trained models while meeting the pre-defined accuracy and privacy constraint of each model\textcolor{black}{, e.g., in Fig.~\ref{fig:deduplication-overview}, the $\epsilon$ of model $M_i$ is bounded by $\epsilon^*_i$}. While the traditional model deduplication problem focuses on addressing the large search space, our identified problem poses new research challenges as follows:

\noindent
\textcolor{black}{
\textbf{Challenge 1. Privacy Constraint.} 
} 
Existing approaches allow blocks from a target model to be replaced by blocks from any other models called base models, causing the composed privacy loss (Sec.~\ref{sec:privacy-budget-derivation}) of the deduplicated model unbounded. Our target problem requires a careful selection of base models to minimize overall storage and privacy costs while meeting privacy and accuracy constraints. However, base model selection is a challenging optimization problem, since it is hard to estimate the compression ratio of deduplicating a pair of base and target models under an accuracy constraint.


\noindent
\textbf{Challenge 2. Time and Privacy Costs of Accuracy Validation.} Existing accuracy-aware model deduplication requires frequent assessment of the accuracy of each modified model in downstream tasks, to ensure that any accuracy decline remains within acceptable limits. This validation process presents a trade-off: frequent accuracy checks will delay the processing and escalate privacy costs for private validation datasets, while infrequent validations lead to more deduplication failures and hurt the compression ratio.

\vspace{3pt}
\noindent
\textbf{Novel privacy and accuracy-aware model deduplication (Fig.~\ref{fig:priv-dedup-overview}).}  
\textcolor{black}{To address Challenge 1, we quantify different types of privacy cost composition caused by deduplication, and formalize the base model selection problem. We develop a greedy strategy that first clusters models based on the similarity of model metadata, e.g., model architecture and training dataset identifier, and then select base models with high quality (measured by privacy loss and storage benefits) to deduplicate the rest of the models in their cluster. (See Sec.~\ref{sec:base-model}.)
}

 To address Challenge 2, when deduplicating a pair of base and target models, we propose a dynamic validation strategy coupled with a gradient-based saliency analysis approach to reduce the required number of accuracy validations. At each step, a dynamic number of blocks are deduplicated to gradually separate non-deduplicable blocks from deduplicable blocks.  Then, we leverage the Sparse Vector Technique (SVT)~\cite{Dwork2014DP} to reduce the privacy loss when private validation data is used. This approach transforms the validation steps into a series of boolean questions about whether accuracy drops exceed a threshold, maintaining a fixed privacy budget until the number of negative answers reaches a predetermined cut-off value. 
 (See Sec.~\ref{sec:main_algorithm} and ~\ref{sec:svt}.)  
 
 The key contributions of our work include:

 \begin{figure*}[t]
\centering
\includegraphics[width=1\textwidth]{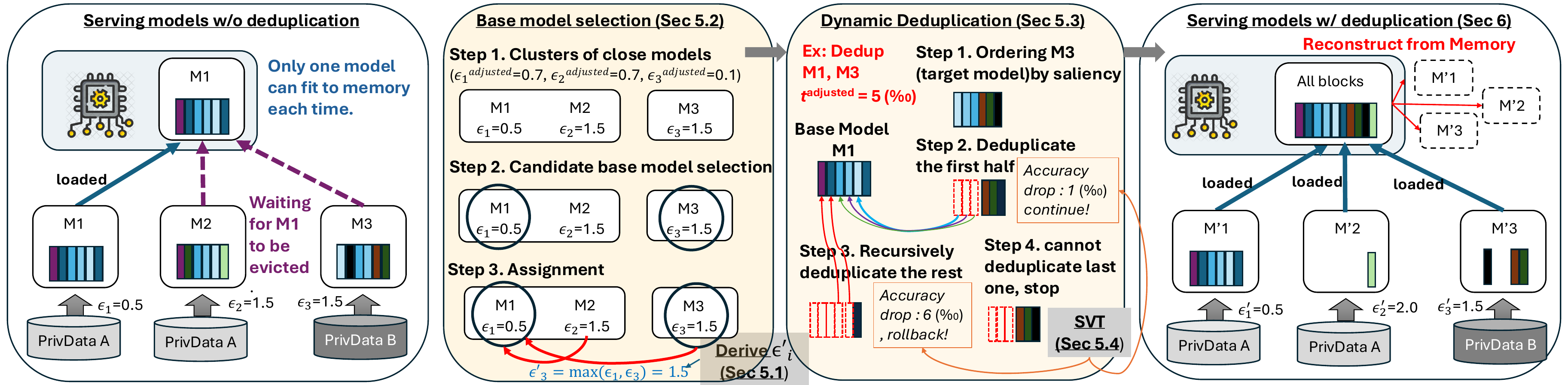}
\caption{ System Overview. In the second box (B2), M1 and M2 are partitioned to the same group because they are trained on the same dataset, while M3 is in a separate group (B2-Step 1). Then, it selects M1 to be the base model, since there is no qualified base model for $M1$ (B2-Step 2). Then, M2 is assigned to M1, while M3, the unused base model in the other group, is also assigned to M1 (B2-Step 3). The next box (B3) shows how M3 deduplicates with its assigned base model M1. M3's blocks are first ordered by saliency ascendingly (B3-Step 1). Then it first deduplicates the left half of the blocks by replacing each block using the most similar block from M1, followed by an accuracy validation. If the accuracy drop is within $0.005$, it recursively deduplicates the right half (B3-Step 2). Otherwise (B3-Step 3), it rolls back the previous step (B3-Step 4), splits the current group into two, and recursively deduplicates the left half. The recursion stops when the number of blocks < 2. }
\label{fig:priv-dedup-overview}
\end{figure*}

\vspace{-7pt}
\begin{enumerate}[wide, labelwidth=!, labelindent=0pt]
    \item We are the first to formalize the problem of model deduplication with privacy constraints, and the privacy loss composition of deduplicated models. 
    \item  We are also the first to design and develop an end-to-end pipeline to optimize and balance privacy, accuracy, and compression ratio, including base model selection, dynamic block deduplication, and SVT-based accuracy validation.
    \item We implement the proposed system 
     and conduct comprehensive evaluations. The results show that, compared to SOTA model deduplication methods that do not consider privacy at all, our approach can improve the compression ratio for deduplicating multiple large language models, vision transformers, and ResNet models, by up to $35\times$ for individual models with privacy costs reduced by up to $3\times$ compared to alternative base model selection designs. The multi-model inference speed is also accelerated by $43\times$ by avoiding swapping models between memory and disk.
\end{enumerate}

%% file: background.tex
\section{Preliminaries}
\label{sec:preliminary}


\noindent
\textbf{Differential Privacy \cite{Dwork2014DP, mcsherry2009privacy}} (DP) is a mathematical framework for quantifying privacy risks in data analysis, and is increasingly adopted in real-world applications by companies like Google~\cite{hard2018federated} and Apple~\cite{ramaswamy2019learning}.

\vspace{-7pt}
\textcolor{black}{
\begin{definition}
[Differential Privacy~\cite{Dwork2014DP}]
A randomized algorithm $\mathcal{M}$ is ($\epsilon$,$\delta$)-differentially private if for all neighboring datasets (i.e. differ by exactly one element) $\mathcal{D}$ and $\mathcal{D'}$, and for $\forall \mathcal{S} \subseteq \mathrm{Range}(\mathcal{M}) $, $Pr[\mathcal{M(D)} \in \mathcal{S}] \leq e^{\epsilon} * Pr[\mathcal{M(D')} \in \mathcal{S}] + \delta$.
$\epsilon$, also called \textbf{privacy budget}, is a metric of privacy loss at a differential change in data. $\delta$ denotes the probability of the privacy guarantee being failed.
\end{definition} 
}
\vspace{-3pt}

\vspace{3pt}

\noindent
Properties of DP. DP is characterized by three key properties. \textit{Sequential Composition} illustrates how privacy costs accumulate when multiple analyses are performed on the same dataset. When analyses are conducted on disjoint subsets of data, \textit{Parallel Composition}~\cite{McSherry2009pquery} considers the maximum privacy cost on each dataset as the privacy cost. \textit{Post-processing} ensures that any data-independent processing of differentially private outputs will not incur additional privacy costs. The formal statements of these properties are:

\begin{thm}[Sequential Composition~\cite{Dwork2014DP}]
\label{thm:sequential}
If an algorithm $\mathcal{M}_1$ is $(\epsilon_1, \delta_1)$-DP and $\mathcal{M}_2$ is $(\epsilon_2, \delta_2)$-DP, then their sequential composition $\mathcal{M}(x) = (\mathcal{M}_1(x), \mathcal{M}_2(x))$ is $(\epsilon_1 + \epsilon_2, \delta_1 + \delta_2)$-DP.
\vspace{-3pt}
\end{thm}

\begin{thm}[Parallel Composition\cite{mcsherry2009privacy}]
\label{thm:parallel}
Let $\mathcal{D}_1, \mathcal{D}_2, \ldots, \mathcal{D}_k$ be disjoint subsets of the database $\mathcal{D}$. If algorithm $\mathcal{M}_i$ is $(\epsilon, \delta)$-DP for each $i \in [1,k]$, then the parallel composition $M(x) = (\mathcal{M}_1(\mathcal{D}_1), \ldots, $ $\mathcal{M}_k(\mathcal{D}_k))$ is ($\max_i \epsilon_i, \max_i \delta_i)$-DP.

\end{thm}

\begin{thm}[Post-processing~\cite{Dwork2014DP}]
\label{thm:post-processing}
If $\mathcal{M}$ is $(\epsilon, \delta)$-DP and $\mathrm{f}$ is any data independent function, then $\mathrm{f} \circ \mathcal{M}$ is also $(\epsilon, \delta)$-DP.
\end{thm}

\textcolor{black}{The post-processing property ensures that the privacy budget of a model will not increase with the number of inferences. In addition, given models trained on a dataset $D$, deduplicating them will not affect the overall privacy budget on $D$.} 

\vspace{3pt}
In the context of DP, sensitivity measures the maximum change in the output of a function when a single data point in the input dataset is modified. Formally,  we have the following definition.

\begin{definition}[Sensitivity\cite{Dwork2014DP}]
For a given function $f$, the sensitivity $\Delta f$ is defined as 
$
\Delta f = \max_{\mathcal{D}, \mathcal{D'}} \| f(\mathcal{D}) - f(\mathcal{D'}) \|
$
where $\mathcal{D}$ and $\mathcal{D'}$ are any two neighboring datasets. 
\end{definition}

\noindent
\textbf{Sparse Vector Technique (SVT) \cite{Dwork2014DP}} is a method in DP for answering a large number of queries while only incurring the privacy cost for those queries that exceed a certain threshold. This is achieved by adding noise to both the threshold and each query output. Let cut-off $c$ denote the maximum number of queries to be answered with result exceeding threshold  before halting, $\Delta$ represent the sensitivity, and $\epsilon$ signify the total privacy budget. The privacy budget is divided into two parts: $\epsilon_1$ for the threshold and $\epsilon_2$ for query results, such that $\epsilon_1 + \epsilon_2 = \epsilon$. The algorithm proceeds as follows: noise is sampled from a distribution $\mathrm{Lap}(\frac{\Delta}{\epsilon_1})$ and added to a predetermined threshold. For each query, noise is sampled from a distribution $\mathrm{Lap}(\frac{2c\Delta}{\epsilon_2})$ and added to the query result. If the noisy query result is less than the noisy threshold, it returns a negative answer and the procedure continues; otherwise, it returns a positive answer and increse the counter. The algorithm halts when the counter reaches $c$. A recommended ratio between the two epsilon values is $\epsilon_1:\epsilon_2=1:(2c)^{2/3}$~\cite{lyu2017understanding}.




\vspace{3pt}
\noindent
\textbf{Differentially Private Stochastic Gradient Decent (DP-SGD) \cite{abadi2016deep}.} 
In DP-SGD, privacy is maintained through two primary modifications to the traditional SGD algorithm: gradient clipping and the addition of noise. Mathematically, the update rule for the model parameters \(\theta\) at step \(t\) in DP-SGD can be expressed as:
\[
\theta_{t+1} \leftarrow \theta_t - \eta \left( \frac{1}{B} \sum_{i=1}^{B} \text{clip}(g_i(\theta_t), C) + \mathcal{N}(0, \sigma^2 I) \right)
\]
where \(\eta\) is the learning rate, \(B\) is the batch size, \(g_i(\theta_t)\) is the gradient of the loss function for the \(i\)-th data point in the mini-batch, \(\text{clip}(g_i(\theta_t), C)\) denotes the gradient clipping operation, and \(\mathcal{N}(0, \sigma^2 I)\) represents the Gaussian noise with variance \(\sigma^2\).

\textcolor{black}{Note in this paper, for simplicity, we focus on $\epsilon$, which dominates the privacy budget. However, all our analysis can be easily extended to $\delta$ by applying the same constraints and computations on $\delta$ as on $\epsilon$ since they follow the same composition rules (Theorem~\ref{thm:sequential}, ~\ref{thm:parallel}, ~\ref{thm:post-processing}).}

%% file: overview.tex
\section{Problem Analysis}
\label{sec:privacy-budget-derivation}

{\color{black}
\textcolor{black}{Here, we will introduce a motivating example, the problems with the SOTA deduplication procedure, and an overview of our solution.}

\vspace{-5pt}
{\color{black}
\subsection{Motivating Example}
\label{sec:motivation}


As shown in Fig.~\ref{fig:deduplication-overview}, a broker of a model marketplace needs to manage models $M_1$, $M_2$, and $M_3$, $M_4$ for Buyer-1 and Buyer-2.  S/he configures $\epsilon^*_i$ (privacy loss increase bound) and $u^*_i$ (utility drop bound) for $M_i$ (e.g., leveraging a privacy provenance system~\cite{zhang2023dprovdb}), and deduplicates these models to minimize storage costs and privacy costs (within $\epsilon^*_i$ and $u^*_i$ bound). As a result, the number of blocks is reduced from $20$ to $8$ (\textbf{O1} in Fig.~\ref{fig:deduplication-overview}\circled{2}) so that more models can be cached in memory, reducing brokers' operational costs. 

Buyer-1 represents a startup MLaaS company, which has an overall $\epsilon$ bound of $5.5$. In Fig.~\ref{fig:deduplication-overview}\circled{3}, s/he purchases deduplicated models $M'_1$ and $M'_2$ with a total $\epsilon$ of $5$, and deploys them in a public cloud for serving two users (e.g., User1 and User2 with $\epsilon$ bounds of $1$ and $4$ respectively). MLaaS operators may strive to meet the users' model serving latency requirements as specified in SLOs. Purchasing deduplicated models will reduce the memory footprint for serving both models from $10$ blocks to $6$ blocks, leading to a significant reduction in model serving latency (\textbf{O2} in Fig.~\ref{fig:deduplication-overview}\circled{4}).

Buyer-2 represents an edge application developer, who has an overall $\epsilon$ bound of $4$. In Fig.~\ref{fig:deduplication-overview}\circled{3}, s/he purchases deduplicated models $M_3$ and $M_4$ for deployment on a smart watch for serving a social robot~\cite{lee2020fast} that performs both object recognition task and question answer task at the same time. Assuming User B's edge device can only hold nine blocks, deduplicating models will reduce the number of required blocks from $10$ to $6$, while maintaining compliance with privacy bounds. It means the deduplication at the broker's side (without significantly impacting the utility) enables the serving of both models in memory, eliminates I/O operations fetching blocks from the disk, and significantly reduces the application response latency (\textbf{O3} in Fig.~\ref{fig:deduplication-overview}\circled{4}). After this purchase, Buyer-2 still has a remaining $\epsilon$ bound of $1$ (i.e., $4-max(\epsilon'_2, \epsilon'_3)=1$). 

}

\subsection{Existing Model Deduplication Procedure}
\label{sec:state-of-art-deduplication}

{\color{black}
The state-of-the-art (SOTA) accuracy-aware model deduplication technique~\cite{zhou2022serving}, termed \textbf{Dedup}, assumes that models arrive in order and the first model will not be deduplicated. Starting from the second model, it will repeat the following steps:

\noindent
\textbf{Step 1.} To deduplicate a target model $M_t$, use all previously arrived models to serve as potential block providers, called base models. 

\noindent
\textbf{Step 2.} Order all blocks from $M_t$ by some saliency measures (e.g., the third quantile of weight values in the block~\cite{zhou2022serving}) ascendingly.

\noindent
\textbf{Step 3.} For each block $b \in M_t$ in the ordered list, run the following sub-steps:
(3-1) Select a block $b'$ that is most similar to $b$ from the collection of blocks in all base models (e.g., using locality sensitive hashing~\cite{zhou2022serving}). 
(3-2) Substitute $b$ with $b'$, resulting in a modified model $M'_t$.
(3-3) Evaluate the accuracy of the modified target model using a validation dataset. (This step could be performed once every $N$ iterations, naively reducing the validation frequency.)
(3-4) If the accuracy drop exceeds a threshold, undo the block replacement and stop early, otherwise (3-5) Move to the next block $b \in M_t$. 

\noindent
Limitations of the SOTA method for our targeting problem include: 

\noindent
(\underline{L1}) No existing works have formalized privacy budget derivation and quantification for the deduplicated models.

\noindent
(\underline{L2}) Step 1 naively takes all available models as base models, without checking whether combining the target model with blocks from all base models will cause a privacy budget increase that violates the privacy budget bound of the target model.

\noindent
(\underline{L3}) Step 2 uses simple saliency measures, which we found less effective for models with noisy parameters trained with DP.

\noindent
(\underline{L4}) Step 3 adopts a static strategy to validate the accuracy of a deduplicated model. (As mentioned, the expensive accuracy validations compose a major bottleneck of the deduplication process.)

\noindent
(\underline{L5}) If the validation dataset used in Step 3-3 is private, the accuracy comparison in Step 3-4 will introduce additional privacy costs not considered by existing approaches.
}


{\color{black}
\vspace{-5pt}
\subsection{Methodology Overview}
As illustrated in Fig.~\ref{fig:priv-dedup-overview}, an overview of our work is as follows.

\noindent
\textbf{Private Budget Derivation (Sec.~\ref{sec:privacy-budget-derivation}).} To address \underline{L1}, \textcolor{black}{we are the first to quantify the privacy loss of a target model $M_t$ deduplicated by a base model $\hat{M}$, while $M_t$ and $\hat{M}$ could be trained on the same, disjoint, and even overlapping datasets.}

\noindent
\textbf{Base Model Selection (Sec.~\ref{sec:base-model}).}  To address \underline{L2}, we first formalize the base model selection as a combined set partitioning and generalized assignment problem, \textcolor{black}{which partitions the models into base models and target models and assigns base models to target models.} \textcolor{black}{However, there is no easy way to efficiently estimate the compression ratio for deduplicating each pair of models without incurring additional privacy costs.} Therefore, we developed a greedy algorithm to prioritize high-quality models \textcolor{black}{that have lower privacy costs and higher storage benefits (i.e., more similar to target models and with fewer qualified base models)} to serve as the base model. 

\noindent
\textbf{Deduplication with Saliency Analysis and Dynamic Validation (Sec.~\ref{sec:main_algorithm}).} 
\underline{L3} and \underline{L4} are correlated. We compared several existing weight saliency measurements~\cite{li2017pruning, liu2021group, sun2024simple, frantar2022gptq} and found the gradient magnitude~\cite{frantar2022gptq} well balances saliency profiling latency and effectiveness. We also propose our dynamic deduplication scheme where dynamic ranges of non-salient blocks are grouped and deduplicated together to avoid accuracy validation failures. 

\noindent
\textbf{Sparse Vector Technique (SVT) for
Validation Using Private Data (Sec.~\ref{sec:svt}).} To address \underline{L5}, we abstract accuracy validation as a boolean question about whether the accuracy drop is above a threshold and apply SVT to provide a fixed budget for a pre-specified number of failed validations (i.e., the boolean questions return negative results). 
}

\vspace{-5pt}
\section{Novel Automatic Deduplication}

\textcolor{black}{In this section, we will present our solutions in detail.}}
\label{sec:deduplication}
\subsection{Privacy Budget Derivation}
\label{sec:privacy-budget-derivation}

The privacy budget of $M'_t$ in the above deduplication process can be determined by Theorem~\ref{privacy-with-deduplication}, of which the intuition is as follows. Consider the case where two similar blocks $b_t$ and $\hat{b}$ are from model $M_t$ with $\epsilon_t$ and $\hat{M}$ with $\hat{\epsilon}$ that are trained on dataset \textcolor{black}{$D_t$ and $\hat{D}$} respectively. Using $\hat{b}$ to replace $b_t$ will change $\epsilon_t$ into $\epsilon'_t$ in three situations: (1) \textbf{Intra-Model ($M_t = \hat{M}$)}: If $b_t$ and $\hat{b}$ are from the same model, they can be safely deduplicated with $\epsilon'_t = \epsilon_t$. 
(2) \textbf{Inter-Model Intra-Data ($M_t \neq \hat{M} \wedge $\textcolor{black}{$D_t = \hat{D}$})}: If $b_t$ and $\hat{b}$ are from different models that were trained with the same dataset, we have $\epsilon'_t = \epsilon_t + \hat{\epsilon}$ by applying DP's sequential composition property~\cite{dwork2008differential, McSherry2009pquery}, detailed in Sec.~\ref{sec:preliminary}. 
(3) \textbf{Inter-Model Inter-Data ($(M_t \neq \hat{M}) \wedge$\textcolor{black}{$ (D_t \cap \hat{D} =\Phi)$})}: If $b_t$ and $\hat{b}$ are from different models that were trained on disjoint datasets,\textcolor{black}{ we have $\epsilon'_t = $\textcolor{black}{$\max(\epsilon_t, \hat{\epsilon})$} by applying DP's parallel composition ~\cite{dwork2008differential, McSherry2009pquery}}, detailed in Sec.~\ref{sec:preliminary}. 
 

\vspace{-2pt}
\begin{thm}\label{privacy-with-deduplication}
Suppose that $d$ disjoint training datasets are used for training all models.
Let \textcolor{black}{$\hat{M}_t$} be a $(\epsilon_t, \delta_t)$-DP model to be deduplicated, and let \textcolor{black}{$\mathcal{\hat{M}}$}$=\{\hat{M}_1, ..., \hat{M}_k\}$ be a set of $k$ distinct differentially private models, each satisfying \textcolor{black}{$(\hat{\epsilon_i},\hat{\delta_i})$-DP}, serving as potential block providers. Any two models in $\{M_t\} \cup \mathcal{\hat{M}}$ are trained on either the same dataset or completely disjoint datasets from each other. Let \textcolor{black}{$\mathcal{\hat{M}}_j=\{\hat{M}_{j_1},...\hat{M}_{j_l}\}$} be the group of models trained on the dataset \textcolor{black}{$D_j$}. Then, on \textcolor{black}{$D_j$},
the resulting deduplicated model, denoted as $M'_t$, satisfies \textcolor{black}{($\epsilon_t \cdot \mathds{1}_{M_t,D_j} +\sum_{i=1}^l  \hat{\epsilon_{j_i}}$, \quad$\delta_t \cdot \mathds{1}_{M_t,D_j}+\sum_{i=1}^l \hat{\delta_{j_i}}$)}-DP, where the indicator function $\mathds{1}_{m,D_j}$ is equal to 1 if model $m$ is trained on \textcolor{black}{$D_j$}, otherwise 0. If $\mathcal{\hat{M}}=\emptyset$, then $M'_t$ satisfies $(\epsilon_t, \delta_t)$-DP.

Given any disjoint \textcolor{black}{$D_j$} ($1\le j\le d$), if the derived model $M'_t$ satisfies ($\epsilon^j, \delta^j$)-DP, the derived model $M'_t$ satisfies ($\max_j \epsilon^j, \max_j \delta^j$)-DP on the union of datasets $\cup_{j=1}^d $$\textcolor{black}{D_j}$.



\end{thm}

\vspace{-5pt}
\noindent
\textbf{Proof.} 
Because any two models are trained on either the same dataset or the datasets completely disjoint from one another, we can divide the models in \textcolor{black}{$\mathcal{\hat{M}}$} into groups by their training datasets.  \textcolor{black}{$\mathcal{\hat{M}}_j=\{\hat{M_{j_1}},...\hat{M_{j_l}}\}$} is a group of models using training data \textcolor{black}{$D_j$}. By DP composability, the collection of blocks from this group of models satisfies (\textcolor{black}{$\sum_{i=1}^l \hat{\epsilon_{j_i}}$}, \textcolor{black}{$\sum_{i=1}^l \hat{\delta_{j_i}}$})-DP. Because the deduplication algorithm iterates over all blocks to decide replacements, by DP post-processing property, the derived model $M'_t$ satisfies (\textcolor{black}{$\sum_{i=1}^l \hat{\epsilon_{j_i}}$}, \textcolor{black}{$\sum_{i=1}^l \hat{\delta_{j_i}}$})-DP on \textcolor{black}{$D_j$} when the original model $M_t$ is not trained on \textcolor{black}{$D_j$}. If $M_t$ is trained on \textcolor{black}{$D_j$}, $M'_t$ satisfies \textcolor{black}{($\epsilon_t+\sum_{i=1}^l \hat{\epsilon_{j_i}}$, $\delta_t+\sum_{i=1}^l \hat{\delta_{j_i}}$)}-DP by DP sequential composition. Given any disjoint dataset \textcolor{black}{$D_j$} ($1\le j\le d$), if the deduplicated model $M'_t$ satisfies ($\epsilon^j, \delta^j$)-DP, it satisfies ($\max_j \epsilon^j, \max_j \delta^j$)-DP on the union of disjoint datasets $\cup_{j=1}^d $\textcolor{black}{$D_j$}, by the parallel composition property~\cite{McSherry2009pquery}.

 \noindent
 \textbf{Generalization.} We can easily remove the disjoint dataset assumption by abstracting the problem as a graph. Assume that for graph $G$, each node represents a base model, and two nodes \textcolor{black}{$\hat{M}_p$} and \textcolor{black}{$\hat{M}_q$} have an edge if and only if their training datasets overlap.  Regarding each connected component $C_j$ of $k$ nodes as a model group \textcolor{black}{$\{\hat{M}_{j_1},...\hat{M}_{j_k}\}$}, similarly, the collection of blocks from it satisfies $\epsilon^j= \sum_{i=1}^{k}\epsilon_{j_i}$ -DP. Suppose $G$ has $d$ connected components, then the derived model $M'$ satisfies ($\max\limits_{1 \le j\le d} \epsilon^j$)-DP.


\noindent


%

\input{basemodel}

\input{algorithm}

\subsection{SVT for Validation Using Private Data}
\label{sec:svt}

Suppose that $f(D, M)$ evaluates the accuracy of a model $M$ on a private validation dataset $D$, which is disjoint from the training set. Given a classification model, its utility $f$ has a sensitivity of $\frac{1}{|D|}$ because the number of correct predictions changes at most by 1 on two neighbor datasets and the accuracy is calculated as the percentage of the correct predictions. Suppose that $M'$ is the new model after a deduplication step on $M$. Then, the utility drop, i.e., $f(D,M)-f(D,M')$ has sensitivity $\frac{2}{|D|}$. Each deduplication step is followed by a validation accuracy query on $D$, and the answer is used to determine whether the deduplication should stop. Accordingly, Sparse Vector Technique (SVT) works by applying Laplace noise and comparing the utility drop with a noisy threshold, as shown in Alg.~\ref{alg:svt-n}. 
When the validation dataset is large, the privacy budget required for deduplication can be significantly smaller, as the sensitivity decreases inversely with the size of the dataset.

{\color{black}
Alg.~\ref{alg:svt-n} employs two distinct privacy parameters: $\epsilon_1$ for noise added to the threshold, $\epsilon_2$ for the validation accuracy comparison. We adopt the recommended ratio of $\epsilon_1:\epsilon_2 = 1:(2c)^{2/3}$ as proposed by Lyu et al.~\cite{lyu2017understanding}. Cut-off $c$ represents the maximum number of times the deduplication process can fail the utility test (i.e., validation), which is pre-specified. 

}


\begin{algorithm}[t]
\caption{\scriptsize $\epsilon$-DP Deduplicate with SVT}
\label{alg:svt-n}
\scriptsize
\begin{algorithmic}[1]
\State Input: private validation dataset $D$; model $M$ to be deduplicated; allowed utility drop threshold $T$; cut-off $c$; privacy budget $\epsilon$ for validation.
\Procedure{SVT}{$M,D,T,c,\epsilon$}
    \State $\epsilon_1=\frac{1}{1+(2c)^{2/3}}\epsilon,\;\;\epsilon_2=\frac{(2c)^{2/3}}{1+(2c)^{2/3}}\epsilon$;
    \State $\hat{T} = T + \mathrm{Lap}(\frac{2}{|D|\epsilon_1})$; \Comment{Add noise to threshold $T$}
    \State $\text{count}=0$;
    \While{true}
    \State Get next set of blocks to be deduplicated $\mathcal{B}=\{b_1,b_2,...,b_l\}$ proposed by the greedy or dynamic deduplication algorithms.
    \If{$\mathcal{B}=\emptyset$} Halt;
    \EndIf
    \State $M'$ = Deduplicate($M,\mathcal{B}$) ;
    \State $\Delta u=f(D,M)-f(D,M')$; \Comment{Compute utility drop}
    \State $v_i = \Delta u + \mathrm{Lap}(\frac{4c}{|D|\epsilon_2})$; \Comment{Add noise to utility drop}
    \If{$v_i \geq \hat{T}$} \Comment{ Deduplication fails}
    \State Output $M$;
    \State $\text{count}=\text{count}+1$;
    \If{$\text{count}\ge c$} Halt;\Comment{Reach cut-off $c$}
    \EndIf
    \Else \Comment{Deduplication succeeds}
    \State Output $M'$;
    \State $M=M'$;
    \EndIf
    \EndWhile
\EndProcedure
\end{algorithmic}
\end{algorithm}
\setlength{\textfloatsep}{3pt}

%% file: basemodel.tex
{\color{black}
\subsection{Base Model Selection}
\label{sec:base-model}
\subsubsection{Problem Analysis and Formalization}
A base model is a block provider, e.g., $M_1$ in Fig.~\ref{fig:deduplication-overview}.

According to the privacy analysis in Sec.~\ref{sec:privacy-budget-derivation}, the more base models used, the more the privacy loss ($\epsilon$) of the target model increases, which is the \textbf{key } to optimize the privacy loss of the target model.  
%
%
%
The research question is how to select base models and assign target models to base models to minimize storage and privacy costs while meeting the accuracy and privacy constraints.  We start from two assumptions: (Assumption-1) each target model can only use one base model for deduplication, which practically reduces the privacy loss of the resulting models and the deduplication overheads. (We allow many target models to share one base model to maximize their utilization) and (Assumption-2) We do not allow a model $\hat{M}$ to be assigned as a base model and a target model at the same time, because deduplicating $\hat{M}$ as a target model may cause parameter and accuracy changes in all models that have been deduplicated using $\hat{M}$ as a base model. 

\vspace{5pt}
\noindent
\textbf{Problem Definition.} Given a set of models trained on the same dataset or disjoint datasets, $\mathcal{M}=\{M_1, ..., M_n\}$, satisfying $\epsilon_1,\dotsc,\epsilon_n$-DP respectively. These models have utilities (accuracy
) $u_1, \dots, u_n$ respectively. Each model is split into (tensor) blocks of equal size. 
Each model $M_j (j = 1, \dots, n)$ has an $\epsilon$ increase threshold $\epsilon^*$ and a utility drop threshold $u_j^*$. The deduplication process must satisfy (1) $M_j$'s $\epsilon$ increase should be bounded by $\epsilon^*$, (2) $M_j$'s utility drop should be bounded by $u_j^*$, and (3) the \textit{fairness rule} ensuring models trained with higher $\epsilon$ on the same dataset must maintain higher accuracy than those trained with lower $\epsilon$. 

We suppose $n_{ij}$ in an $n\times n$ matrix $\mathcal{N}=\{n_{ij}\}$ represents the number of blocks from $M_j$ (as a target model) that can be replaced by blocks from $M_i$ (as a base model) with $M_j$'s utility drop bounded by $u_j^*$. 
$\mathcal{A}=\{a_{ij}\}$ represents the partitioning and assignment matrix that partitions models into target models and base models and assigns target models to base models. $a_{ij} = 1$ means $M_j$ is assigned to the base model $M_i$. 
%
The objective function and constraints are formalized below, where $f_{\epsilon}(M_i, M_j)$ denotes the privacy budget ($\epsilon'_i$) of the target model $M_j$ using $M_i$ as a base model based on the derivation in Sec.~\ref{sec:privacy-budget-derivation}.
$\lambda$ is a regularization factor to balance the two objectives (1) To maximize the total number of blocks that can be deduplicated (replaced) ($\sum\limits_{i=1}^{n}\sum\limits_{j=1}^{n} (n_{ij})$) and (2) to minimize the target models' overall $\epsilon$ increase caused by deduplication 
 ($\sum\limits_{i=1}^{n}\sum\limits_{j=1}^{n} (f_{\epsilon}(M_i, M_j))$). The optimization process must satisfy the privacy constraint $\epsilon^{adjust}_j$ (Eq.~\ref{eq:problem_constraint1}) explained later, the Assumption-1 (Eq.~\ref{eq:problem_constraint2}), and Assumption-2 (Eq.~\ref{eq:problem_constraint3}). 

\vspace{-5pt}
\begin{equation}
\label{eq:basemodel-selection}
\underset{a_{ij=0,1}(1 \leq i,j \leq n)}{\max} \sum\limits_{i=1}^{n}\sum\limits_{j=1}^{n} (n_{ij}-\lambda \times f_{\epsilon}(M_i, M_j))\times{a_{ij}}, \quad s.t.
\end{equation}
\begin{equation}
\label{eq:problem_constraint1}
f_\epsilon(M_i, M_j)-\epsilon_j < \epsilon^{adjusted}_j \qquad\text{if}\quad a_{ij}=1
\end{equation}
\begin{equation}
\label{eq:problem_constraint2}
\sum\limits_{i=1}^{n}a_{ij}=1
\end{equation}
\begin{equation}
\label{eq:problem_constraint3}
\forall M_i, \nexists M_k, M_j (i \neq k \neq j) \quad\text{satisfying}\quad a_{ik}=1 \wedge a_{ji}=1
\end{equation}

\noindent
\textbf{$u^{adjust}_i$ (used for computing $n_{ij}$) and $\epsilon^{adjust}_j$:} To meet the fairness rule, we enforce that the orderings of these models by the $\epsilon$ and by $utility$ do not change after deduplication. Then, if all models satisfy the fairness rule before deduplication, these models still adhere to the fairness rule after deduplication. Suppose $M_i$ and $M_{i-1}$ are two consecutive models in the sequence of $k$ models trained on the same dataset ordered ascendingly by utility ($u_i > u_{i-1}$), and if the user-specified utility ranges of the deduplicated models $M'_{i}$ and $M'_{i-1}$ are [$u_i-u^*_i$, $u_i$] and [$u_{i-1}-u^*_{i-1}$, $u_{i-1}$] and they overlap, 
$u'_i < u'_{i-1}$ is allowed
, which violates the order by utilities. To eliminate such overlapping ranges, we adjust $u_i^*$ to be $u_i^{adjusted}=min(u_i^*, u_{i}-u_{i-1})$ for $i=2,\dots,k$ so that the deduplication algorithm always ensures $u'_i > u'_{i-1}$. Similarly, we set $\epsilon^*_i$ to be $\epsilon_i^{adjusted}=min(\epsilon_i^*, \epsilon_{i}-\epsilon_{i-1})$ for $i=2,\dots,k$.

\noindent
\textbf{Analysis.} If we fix a subset of models $\mathcal{B}\subset \mathcal{M}$ to be base models, 
the problem is a generalized assignment problem (GAP)~\cite{cohen2006efficient}, where one or more target models (i.e., tasks in the classical GAP problem) are assigned to each base model (i.e., agents in the classical GAP). The GAP problem was proven to be NP-hard~\cite{cohen2006efficient, cattrysse1994set}. However, instead of having a fixed $\mathcal{B}$, our problem additionally requires searching for a set-partitioning scheme that divides $\mathcal{M}$ into $\mathcal{B}$ and $\mathcal{M}-\mathcal{B}$.

Moreover, the problem is a black-box optimization problem~\cite{doerr2011black}, since it is expensive to estimate $n_{ij}$. Performing actual deduplication to determine $n_{ij}$ is generally expensive. Using weight-level similarity to estimate $n_{ij}$ is also impractical since the similarity function will compose $\epsilon$ of input models. Therefore, we developed a greedy strategy to address the problems.

\vspace{-3pt}
\subsubsection{A Greedy Strategy.}
\label{sec:greedy}
The main intuition is that the quality of the base models is more important than the quantity. A high-quality base model should satisfy the following requirements: (\textbf{R1}) \textbf{Similar to target models.} Using a base model with greater similarity in model architecture, training dataset, utility, and privacy budget to the target model usually leads to a better compression ratio (with the same utility constraint). (\textbf{R2}) \textbf{Low privacy budget.} Usually, a base model with a lower privacy loss leads to a smaller increase in privacy cost in the duplication model. (\textbf{R3}) \textbf{Low chance to be deduplicated (Low opportunity cost)}. A model having fewer qualified base models than qualified target models is more suitable to serve as a base model than a target model. 
Based on the intuition, our key idea is to cluster models by similarity of model metadata, and in each cluster, we select the models satisfying R1, R2, and R3, to serve as base models for the rest of the cluster:

\begin{enumerate}[wide, labelwidth=!, labelindent=0pt]
    \item \textbf{Partitioning Models Based on Similarity.} We first cluster the models from $\mathcal{M}$ based on public metadata, such as model architecture,  public training dataset, the anonymous identifier of the private fine-tuning dataset. We used a simple hierarchical clustering strategy and ensured that all models in the same cluster have the same model architecture and similar datasets.
    \item \textbf{Assigning Dangling Models as Base Models.} We define dangling models to be models that do not have any qualified base models in their clusters (i.e., $M_j$ cannot find any $M_i$ satisfying Eq.~\ref{eq:problem_constraint1}). There must be at least one dangling model in each cluster--the model with the smallest $\epsilon$. We only allow these dangling models to serve as the base models to deduplicate other models in the cluster since they are similar to other models (i.e., can deduplicate a good number of blocks from other models) and zero blocks can be deduplicated from them if assigned as target models, thus maximizing the storage benefits in its local cluster. 
The process is described in Alg.~\ref{alg:dangling-models}. 
    \item \textbf{Intra-Group and Inter-Group Base Model Selection.} For each non-dangling model, it must be able to find at least one qualified base model in its cluster. If it has two or more qualified base models in its cluster, selecting any of them may lead to similar compression ratios according to our observation. Therefore, we will select the one that brings minimum privacy cost increase as its base model (line $6$-$9$ in Alg.~\ref{alg:base-model-main}). Then, for those dangling models that are never used (i.e., $count==0$), we search for qualified base models from other clusters (line $10$-$15$ in Alg.~\ref{alg:base-model-main}). 
\end{enumerate}

\setlength{\textfloatsep}{0pt}
\begin{algorithm}[h]
\caption{\scriptsize Identify dangling models in a group $\mathcal{G}=\{M_1, ..., M_k\}$}
\label{alg:dangling-models}
\scriptsize
\begin{algorithmic}[1]
\Procedure{CandidateBaseModelGen}{$\mathcal{G}$}
\State  $\mathcal{B}\leftarrow \{\}$; $\mathcal{T}\leftarrow \mathcal{G}$
\For{$M_i$ in $\mathcal{T}$} \Comment{Identify dangling models and add them to $\mathcal{B}$}
    \If{$\epsilon_j \geq \epsilon^{adjusted}_i$ for all $j\in\{1,...,k\}$ and $j\neq i$}
        \State $\mathcal{B} = \mathcal{B} \cup \{M_i\}$; $\mathcal{T} = \mathcal{T} - \{M_i\}$;
    \EndIf
\EndFor
\State \Return
\EndProcedure
\end{algorithmic}
\end{algorithm}

\vspace{-10pt}

\setlength{\textfloatsep}{0pt}
\begin{algorithm}[h]
\caption{\scriptsize Select a base model $\hat{M}$ for a target model $M_t$ in a Group $\mathcal{G}$}
\label{alg:base-model-step}
\scriptsize
\begin{algorithmic}[1]
\Procedure{BaseModelSelectStep}{$M_t$, $\mathcal{G}$}
\State $\mathcal{B} \leftarrow $ The set of candidate base models in $\mathcal{G}$; 
\State Find base model $\hat{M} \in \mathcal{B}$ with the smallest $\hat{\epsilon}$ to satisfy $f_{\epsilon}(\hat{M}, M_t)-\epsilon_t < \epsilon^{adjusted}_t$ to serve as the base model for $M_t$; $\hat{M}.count++$ 
\State \Return $\hat{M}$
\EndProcedure
\end{algorithmic}
\end{algorithm}

\vspace{-10pt}
\setlength{\textfloatsep}{0pt}
\begin{algorithm}[h]
\caption{\scriptsize Base model selection for all models in $\mathcal{M}$}
\label{alg:base-model-main}
\scriptsize
\begin{algorithmic}[1]
\Procedure{BaseModelSelectMain}{$\mathcal{M}$}
\State $\mathcal{C}=\{\mathcal{G}_1, ..., \mathcal{G}_l\} \leftarrow $ $Cluster(\mathcal{M})$
\For{$\mathcal{G}_i \in  \mathcal{C}$}
 {$CandidateBaseModelGen(\mathcal{G}_i)$}
\EndFor
\For{$\mathcal{G}_i \in  \mathcal{C}$}
\State sort ($\mathcal{C}' = \mathcal{C}-\mathcal{G}_i$) based on similarity to $\mathcal{G}_i$
\For{$M_j \in \mathcal{T}_i$}
\State $ret \leftarrow BaseModelSelectStep(M_j, \mathcal{G}_i)$
\State record that $M_j$'s base model is $ret$
\EndFor
\For{$M_k \in \mathcal{B}_i$}
\If{$M_k.count == 0$}
\While{$ret = NULL \wedge next\_cluster(\mathcal{C}') \neq NULL$}
\State $\mathcal{G}_k \leftarrow$ $next\_cluster(\mathcal{C}')$
\State $ret \leftarrow BaseModelSelectStep(M_k, \mathcal{G}_k)$
\EndWhile
\State record that $M_k$'s base model is $ret$
\EndIf
\EndFor
\EndFor
\Return
\EndProcedure
\end{algorithmic}
\end{algorithm}

%% file: algorithm.tex
\vspace{-10pt}
\subsection{Deduplication Algorithms}
\label{sec:main_algorithm}

\subsubsection{Deduplication Problem Reduction}
\label{sec:two-model-deduplication}
 Once we select the base model for each target model, the privacy budget increase becomes a constant and the privacy and fairness constraints are ensured. Then, the model deduplication problem is reduced to a set of two-model deduplication problems. Each problem involves a target model $M_j$ and a base model $M_i$. The optimization objective is to maximize \textcolor{black}{$n_{ij}=|M_i\cap M'_j|$}, i.e., the number of unique blocks in the deduplicated model $M'_j$ that are from $M_i$ (i.e., minimize the compression ratio) while meeting the utility drop constraint formalized below.
\begin{gather*}
    \max {n_{ij}} \text{ with } \\
    \textcolor{black}{M'_j} = \text{deduplicate}(M_j; M_i \text{ as base model}) \;s.t. \\
    u_j-u'_j < \textcolor{blue}{u}^{adjusted}_j 
\end{gather*}






\subsubsection{Problem Analysis}
%
Given any block from $M_j$, it can be replaced by any block from $M_i$ (or no replacement), leading to \textcolor{black}{$(|M_i|+1)^{|M_j|}$} possible deduplication plans. Depending on the block size, deep learning models in our experiments may have hundreds to thousands of blocks.
Therefore, it is computationally inhibiting to exhaustively search and evaluate the storage costs and accuracy drop of all deduplication plans. 
%
%
Following the existing deduplication framework presented in Sec.~\ref{sec:state-of-art-deduplication} to perform two-model deduplication, it performs a validation after deduplicating every $N$ blocks. When $N$ is set to a small number, the frequent utility validation operations will bottleneck the deduplication process. However, if we set $N$ to a large number, it often deduplicates too many blocks to keep the utility drop within the constraint, due to the lack of sophisticated saliency analysis to order the blocks. A poor saliency measurement may mix a few salient blocks, which will significantly impact the model's utility if deduplicated, with many non-salient blocks in one batch. Then, the salient blocks will cause a validation failure, which will further cause \textit{all} deduplications in the batch, including the deduplications of non-salient blocks, to be rolled back. This will lead to missed opportunities and wasting of resources. However, saliency analysis is a challenging task in nature and made even more difficult by the noises introduced by DP~\cite{rust2023differential}.





\vspace{-5pt}

\subsubsection{Our Dynamic Deduplication Algorithms}
We resolve the problem in two steps: (1) We carefully examine and compare the block saliency measurements for our target problem.
(2) We design new algorithms that dynamically adjust the size of each group to balance the frequency of validations and the number of rollbacks (validation failures). For example, at the beginning, a large group of the least salient blocks should be deduplicated together. As we progress through the ordered list, the group size gradually decreases, ensuring that non-salient blocks are checked in a fine-grained manner.

\vspace{3pt}
\noindent
\textbf{Block saliency measures.} 
Weights that significantly impact the prediction are called ``salient weights''~\cite{huang2023elastictrainer,sun2024simple}. There are many ways to measure the saliency of weights for model explanability~\cite{Simonyan2013DeepIC}, compression~\cite{Lee2020A,li2017pruning,frantar2022optimal,frantar2022gptq}. The most popular ones include the weight magnitude~\cite{zhou2022serving}, activation magnitude~\cite{lin2023awq}, and the Fisher information that involves the square of gradients~\cite{liu2021group}. We find that the Fisher information metric is the most accurate among them. However, it is also the most computationally intensive. In addition, the squared gradients are usually very small and may introduces unstable numerical results. Therefore, we adopted the absolute value of gradient (called gradient magnitude) as our saliency measure. {\textcolor{black}{The saliency of a block is defined as the mean of the gradient magnitude of each weight within a block, which is obtained by one forward and backward pass of a dataset. This dataset could be a public dataset or a disjoint partition from the private training/validation dataset. In the latter case, we added noise to the gradient, and this one-time privacy cost is usually smaller than the model's $\epsilon$ and thus gets absorbed due to DP's parallel composition.}}


\begin{algorithm}[t]
\caption{\scriptsize Dynamic-Range Deduplication (\texttt{DRD})}
\label{alg:dynamic}
\scriptsize
\begin{algorithmic}[1]
\State Input: Target and base models represented by a list of block identifiers $B$, $\hat{B}$; utility drop threshold $T$; minimum \#blocks in a batch $L$.
\Procedure{Deduplication}{$B,\hat{B},T$}
\State Initialize $l=0;r=|B|-1$;
\State Recursion($B,\hat{B}, l,r,T$);
\State \Return;
\EndProcedure

\Procedure{Recursion}{$B,\hat{B}, l,r,T$}
\If{$r-l <L$} \Comment{Base case, skip a small batch}
 \State \Return;
\EndIf
\State $m=(r+l)//2$;
\For{block j in $l,l+1,\dots,m$} \Comment{Try deduplicate the left half}
\State Replace block $B[j]$ by the most similar block in $B \cup \hat{B}$; 
\EndFor
\State Evaluate model on validation set, compute utility drop $\Delta u$;
\If{$\Delta u >=T$} \Comment{If validation fails}
\State Rollback the deduplication of block $l,l+1,\dots,m$;
\If{$l<r$}
    \State Recursion($B,\hat{B},l,m,T$); \Comment{Recursively deduplicate the range}
\EndIf
\EndIf
\State Recursion($B,\hat{B}, m+1,r,T$); \Comment{Recursively deduplicate the right half}
\State \Return;
\EndProcedure
\end{algorithmic}
\end{algorithm}
\setlength{\textfloatsep}{3pt}

\vspace{3pt}
\noindent
\textbf{Group deduplication algorithms.} 
%
Our idea is to dynamically determine the group size and re-evaluate some blocks in a group that fail to be deduplicated. We designed a recursive dynamic-range deduplication (\texttt{DRD}) algorithm, as formalized in Alg. ~\ref{alg:dynamic}.
{\color{black}
 %
%
\texttt{DRD} begins by dividing the input of blocks (i.e., input range) into two halves. It then attempts to deduplicate the left half. If this deduplication fails (i.e., the utility drop exceeds the bound), the algorithm recursively runs on the left half (Line 14) so that at every recursion level, the number of blocks that are deduplicated in a batch will be reduced by half. After deduplicating the left half, the algorithm recursively deduplicates the right half (Line 15). The base case for this recursion occurs when the input range has fewer than $L$ blocks, and it skips such a small range because such remaining blocks are likely to be salient blocks.
We also proposed a variant of \texttt{DRD}, called Dynamic-Range-Expansion Deduplication (\texttt{DRED}), slightly different from \texttt{DRD} in Line 17. \texttt{DRED} will recursively deduplicate the blocks from $m+1$ to $|B+1|$, rather than $m+1$ to $r$ (i.e., changing line 17 to Recursion($B, \hat{B}, m+1, |B|-1, T$)). It allows for more aggressive (coarser-grained) deduplication attempts.
}

Given a block $b$ in each deduplication batch, a block from the base model most similar to block $b$ is chosen to replace $b$. To measure similarity, we compared several distance measures, such as \texttt{l1-norm}, \texttt{l2-norm}, and \texttt{cosine} and chose to directly use \texttt{l2-norm} (i.e., Euclidean distance). We use pairwise comparison for best accuracy, since it is not the bottleneck of our target scenario (less than $10\%$ of overheads). If needed, it can be easily replaced by a faster but more error-prone locality-sensitive hashing~\cite{datar2004locality, mao2017s2jsd} technique.

%% file: implementation.tex
\subsection{Implementation}
\label{sec:storage}
A model may contain many parameter tensors, e.g., embedding vectors, weight matrices, or convolutional filters. In our implementation, each parameter tensor is flattened into a one-dimensional array and then partitioned into blocks with fixed size. The last block will be padded with zeros. Some parameter tensors, such as biases and layer norms, are much smaller than the block size, so they are not partitioned and are managed separately. 

A deduplicated model is stored in three parts
: (1) A 2-D array that contains the blocks for all models, where each row represents a block flattened into a 1-D array. (2) A dictionary for each model containing tensor shapes and extra weights, such as biases and norms that are much smaller than the block size and are handled separately. 
(3) A list of block indices for each model. For example, a model with $K$ blocks 
 would have a $K$-way array, with the $i$-th element specifying the $i$-th block's row index in the 2-D array.


Serving a deduplicated model requires reconstructing the model from the blocks and the dictionaries. 
%
%
Whenever a new query requires a model that is not reconstructed, if there is insufficient memory, a victim model will be evicted following the Least Recently Used (LRU) policy to free up memory space for reconstructing the new model. The system then retrieves the model ID, looks up the corresponding model constitution and extra weights, and reconstructs the queried model. The process is accelerated by reusing a reconstructed model with the same architecture and leveraging our array-based indexing to access blocks for replacement. 

%% file: experiments.tex
\section{Experiments}
\label{sec:experiments}
\textcolor{black}{We conducted comprehensive evaluations with an ablation study to demonstrate the effectiveness of our proposed methods.}

\subsection{Experimental Settings}
\subsubsection{Workloads}
To demonstrate the broad applicability of our proposed approaches, we experimented with a diverse range of model architectures and tasks\footnote{\textcolor{black}{Most of the models in this work are obtained by fine-tuning a pre-trained model on (private) datasets. We focuses on full-finetuning (FFT) that will update all model parameters during the fine-tuning process.  The accuracy of FFT outperformed parameter-efficient fine-tuning (PFT) such as Low-Rank Adaptation (LoRA)~\cite{hu2021lora} for scenarios involving large-scale fine-tuning data, low privacy budget, and complex tasks~\cite{biderman2024lora, budifferentially, hu2021lora}. Our approach can also be applied to deduplicate heterogeneous LLM models or LLMs with LoRA weights merged.}}:

\begin{enumerate}[wide, labelwidth=!, labelindent=0pt]
    \item Roberta-Base~\cite{liu2019roberta} (called Roberta): A Transformer encoder for natural language inference tasks, composed of embedding layers, attention layers, normalization layers, and linear layers. It contains $277$ blocks with a block size of $58,982$ floating points.
    \item Vision Transformer-Large (called ViT)~\cite{dosovitskiy2021an}: A Transformer model for image classification tasks, consisting of attention layers, normalization layers, and linear layers. It has $288$ blocks with a block size of $1,048,576$ floating points.
    \item ResNet152~\cite{he2016deep} (Called ResNet): A convolutional neural network with residual connections for multi-attribute classification tasks. It has $238$ blocks with a size of $262,144$ floating points.
\end{enumerate}

These models encompass all major layer types in modern deep learning architectures, ensuring that our proposed algorithms are applicable to a wide variety of models beyond those explicitly tested.
The datasets were pre-partitioned into training, validation, and test sets, which are disjoint with each other. 

The \textit{choice of privacy budgets} is contingent upon the data and model architecture. Existing works employ varied ranges. For instance, [1.0, 8.0] and (0, 2] for both natural language and image classification tasks~\cite{tramer2021differentially,bu2022differentially,yu2021large,ghazi2021deep, fu2023dpsur,nanayakkara2023chances},  [0.08,4.6] for image and tabular data classification tasks~\cite{bernau2021quantifying}, and [1,16] for image processing tasks~\cite{jagielski2020auditing}. 
In our work, we adopted ranges (0.0,10.0] for the Roberta-base models and (0, 5.5] for ViT and ResNet, respectively. 

\textcolor{black}{Our main experiments involve eight scenarios of models shown as A1-A5 and B1-B3 in Tab.~\ref{tab:scenarios}, designed to be diverse and representative. Each scenario involves one cluster of models. The clusters differ in the numbers of ($5-20$) models, model architectures, training datasets, distribution of epsilons, privacy loss increase threshold $\epsilon^*$, and utility drop threshold $\textcolor{black}{u}^*$. (The latter two constraints are shared by all models in each cluster for simplicity of experiment settings.) Following our base model selection algorithm proposed in Sec.~\ref{sec:base-model}, the first model is selected as the base model for all scenarios except A2. All models in A2 are dangling models, and they will use the model from A3 with $\epsilon=0.2$ as their base model.}
\textcolor{black}{Scenarios C1-C4 in Tab.~\ref{tab:scenarios} involve subsets of A1, B2, B3, and A3 respectively, for an ablation study of the base model selection algorithm. 
}

\vspace{3pt}
\noindent
\textcolor{black}{
\textbf{Two Realistic Scenarios.} We also introduce two realistic scenarios corresponding to two types of buyers in Fig.~\ref{fig:deduplication-overview}. 
\begin{enumerate}[wide, labelwidth=!, labelindent=0pt]
\item \textit{Multi-User Many-Model Scenario,} where a broker trained a total of $50$ distinct models, which is a union of models in Tab.~\ref{tab:scenarios}. These models are sold to a buyer who serves those models to multiple independent users (with different trust and payment levels) concurrently on a cloud. (A similar scenario is to directly sell each model to an independent buyer.)
\item \textit{Single-User Heterogeneous-Model Scenario,} where a broker deduplicated seven models to save operational costs, including four ViT models finetuned on CIFAR10 for object detection~\cite{Krizhevsky09learningmultiple}, CelebA (for face attribution classification)~\cite{liu2015faceattributes}, GTSRB (for traffic sign recognition)~\cite{stallkampManVsComputer2012}, and SVHN (for streetview house number recognition)~\cite{netzer2011reading}, respectively, all with $\epsilon=2$, and three RoBERTa-base models finetuned on SST2 (for sentiment analysis)~\cite{socher2013recursive}, IMDB (for review classification)~\cite{maas2011learning}, and QNLI (for question-answering natural language inference)~\cite{wang2018glue}, respectively, all with $\epsilon=5$. It chose to use the first ViT model as the base model to deduplicate the rest six models. Then, a buyer purchased the six deduplicated models and deployed them in a resource-constrained environment to perform various tasks required by a social robot. 
\end{enumerate}
}


\begin{table}[h]
\centering
\scriptsize
\caption{\label{tab:scenarios} {The Deduplication Scenarios }}
\begin{tabular}{|c|c|c|p{3.3cm}|c|c|} \hline
&Model&Data&Epsilons&$\epsilon^*$&$\textcolor{black}{u}^*$\\ \hline \hline
\multicolumn{6}{|c|}{\textcolor{black}{Diverse scenarios for algorithm evaluation and ablation studies (Sec.~\ref{sec:diverse} and Sec.~\ref{sec:exp-ablation})}} \\\hline
A1&Roberta&QNLI~\cite{wang2018glue}&[1.0, 2.0, 4.0, 6.0, 7.0]&1.0&0.015 \\ \hline
A2&Roberta&SST2~\cite{socher2013recursive}&[0.3, 0.4, 0.6, 0.8, 1.0]&0.05&0.015 \\ 
\hline
A3&Roberta&5 MNLI~\cite{williams2018broad} parts&[0.2, 0.4, 0.8, 1.6, 2.0]&0.2&0.015 \\ \hline
A4&ViT&CIFAR100~\cite{krizhevsky2009learning}&[0.5, 0.6, 0.75, 1.0, 2.0]&0.5&0.020 \\ \hline
A5&ResNet&CelebA~\cite{liu2015faceattributes}&[0.4, 0.6, 0.8, 1.0, 2.0]&0.5&0.020  \\ \hline
B1&Roberta&QNLI&[1.0, 2.0, 3.0, 4.0, 5.0, 6.0, 7.0, 8.0, 9.0, 10.0]&1.0&0.015 \\ \hline
B2&ViT&CIFAR100&[0.5, 0.55, 0.6, 0.65, 0.7, 0.75, 0.8, 0.85, 0.9, 0.95]&0.5&0.020 \\ \hline
B3&ResNet&CelebA&[0.2, 0.3, 0.4, 0.5, 0.6, 0.7, 0.8, 0.9, 1.0, 1.1, 1.2, 1.3, 1.4, 1.5, 1.6, 1.7, 1.8, 1.9, 2, 2.1]&0.2&0.020  \\ \hline
\multicolumn{6}{|c|}{\textcolor{black}{Sub-scenarios for base model selection ablation study (Fig.~\ref{fig:bms_ablation})} }\\\hline
C1&Roberta&QNLI&[5.0, 6.0, 7.0, 8.0, 9.0]&5.0&0.015 \\ \hline
C2&ViT&CIFAR100&[0.5, 0.6, 0.75, 1.0, 2.0]&0.5&0.020 \\ \hline
C3&ResNet&CelebA&[0.7 0.8, 0.9, 1.0, 2.0]&0.7&0.020  \\ \hline
C4&Roberta&4 MNLI parts&[0.4, 0.8, 1.6, 2.0]&0.3&0.015  \\ \hline
\end{tabular}
\end{table}

\vspace{-5pt}
\subsubsection{Comparison}
None of the existing model deduplication mechanisms has considered privacy. Therefore, \textbf{to evaluate the effectiveness of our overall deduplication and accuracy validation approach}, we considered 
the following algorithms, which used our base model selection mechanism to ensure privacy. Unless explicitly noted, they all use gradient magnitude for saliency measurement and L2 (Euclidean) distance as similarity measurement:

\begin{enumerate}[wide, labelwidth=!, labelindent=0pt]
    \item Our \textbf{DRD} as formalized in Alg.~\ref{alg:dynamic} and its variant, \textbf{DRED}. 
    \item The SOTA \textbf{Dedup}
~\cite{zhou2022serving} algorithm as described in Sec.~\ref{sec:state-of-art-deduplication}. 
    \item \textbf{Mistique}~\cite{vartak2018mistique} uses MinHash~\cite{broder1997resemblance} to identify and deduplicate similar model weights without considering accuracy. We improve it to use Dedup's saliency measurement and accuracy validation steps described in Sec.~\ref{sec:state-of-art-deduplication} to provide an accuracy guarantee.
    \item The \textbf{Greedy-N} algorithm, which we developed as a static variant of \textbf{DRD} and \textbf{DRED}. Unlike Dedup, it does not stop after a failure of accuracy validation, instead, it will roll back and then move on to process the next batch. 
    \item We also developed a \textbf{Monte Carlo Tree Search (MCTS)}~\cite{chaslot2010monte} algorithm, where each state represents a deduplication scheme, and an action selects a group of $N$ blocks from the target model to be deduplicated using the most similar blocks from the base model and run an accuracy validation. If the validation fails, the reward (i.e., storage costs saving) is computed and back-propagated. It repeats the process until a time budget is achieved.
    
\end{enumerate}

In addition, we also considered two reference baselines.

\begin{enumerate}[wide, labelwidth=!, labelindent=0pt]
    \item \textbf{Original}, which does not perform any deduplication. 
    \item \textbf{Retrain}, which simply finetunes the models using DP-SGD with the new privacy budget achieved by our deduplication approach. This approach provides the upper-bound accuracy with the same privacy budget increase without deduplication.
\end{enumerate}

\textcolor{black} {\textbf{To evaluate our base model selection method}, we extended Alg.~\ref{alg:base-model-step} to allow multiple base models for one target model by iteratively adding qualified base models until the $\epsilon$ increase exceeds the limit, called \textbf{Multi}. We further extend \textbf{Multi} to allow a deduplicated target model to serve as a base model, called \textbf{Cross}. 
}
\vspace{-5pt}

\subsubsection{Measurements.} We considered the following measurements.

\noindent
\textbf{Compression ratio (C.R.).} We measure the compression ratio for both individual models and clusters of models. Let $M_1, ..., M_n$ denote a cluster of models including the base model, and $M'_1, ..., M'_n$represent these models after deduplication. Each model $M_i$ or $M'_i$ represents a set of unique blocks, and $|M_i|$ or $|M'_i|$ represents the number of blocks in the set. The overall compression ratio is computed as $|M'_1 \cup ...\cup M'_n|/|M_1 \cup ... \cup M_n|$. For a deduplicated model $M'_j$ with its base model $M_i$ , the compression ratio is defined as $|M'_j-M_i|/|M'_j|$. \textit{The smaller the compression ratios, the better.}

\noindent
\textbf{Accuracy.}  We measure the actual accuracy of individual models. In the ablation studies, we use $\max(\Delta u)$ to represent the maximal accuracy drop of individual models in the group.

\noindent
\textbf{Model inference latency.} A great benefit brought by the reduction of the memory footprint of models is the saving of the latency for serving these models in resource-constrained environments where not all models can fit into memory, which is measured in seconds.

\noindent
\textbf{Privacy Budget.} The privacy budget for fine-tuning a model on a dataset using DP-SGD contains both $\epsilon$ and $\delta$. This paper focuses on meeting the constraints on $\epsilon$, which dominates the privacy budget. These delta values are sufficiently small, set to the inverse of the training dataset size, ranging from $6.1 \times 10^{-6}$ (CelebA) to $2.0 \times 10^{-5}$ (CIFAR100). For deduplicated models, the epsilon and delta values are computed according to Theorem~\ref{privacy-with-deduplication}. 

\vspace{-5pt}
\subsubsection{Experimental Environments}
\label{sec:environment}
The deduplication experiments are run on a machine with an Intel(R) Xeon(R) Silver 4310 CPU (2.10 Hz) with 128 GB memory and one NVIDIA A10 GPU with 24 GB GPU memory. \textcolor{black}{To simulate a resource-constrained environment, the model serving experiment for the multi-user many-model scenario is run on a c5a.2xlarge instance (8 CPU cores, 16 GB main memory). All other model-serving experiments are run on a c5a.xlarge instance (4 CPU cores, 8 GB main memory)}.

\vspace{-3pt}
\subsection{Evaluation on Diverse Scenarios}
\label{sec:diverse}

\begin{table*}[]
\begin{center}
\scriptsize
\caption{\textcolor{black}{Comparison of Different Deduplication Algorithms}}
\label{tab:tasks}
\scriptsize
{\color{black}
\begin{tabular}{|c||c|c||c|c||c|c||c|c||c|c||c|c||c|c||c|c|}
\hline
 &\multicolumn{2}{|c||}{A1} &\multicolumn{2}{|c||}{A2} &\multicolumn{2}{|c||}{A3}
 &\multicolumn{2}{|c||}{A4} 
 &\multicolumn{2}{|c||}{A5} 
 &\multicolumn{2}{|c||}{B1} &\multicolumn{2}{|c||}{B2} &\multicolumn{2}{|c|}{B3}\\
&C.R.(\%)&\#Val.&C.R.(\%)&\#Val.&C.R.(\%)&\#Val.&C.R.(\%)&\#Val.&C.R.(\%)&\#Val.&C.R.(\%)&\#Val.&C.R.(\%)&\#Val.&C.R.(\%)&\#Val.\\ 
\hline
\hline
Dedup-20 &48.1&42 &95.7&18&\textbf{56.4}&38&\textbf{32.1}&60&\textbf{87.4}&12&40.8&98&34.5&116&85.4&57\\
Dedup-30 &48.1&\textbf{30}&95.7&\textbf{17}&56.9&\textbf{28}&33.4&\textbf{43}&94.9&\textbf{9}&\textbf{39.7}&\textbf{70}&\textbf{32.6}&\textbf{85}&\textbf{85.2}&57\\
\hline
Mistique-20 &84.4&19&100.0&10&\textbf{77.3}&26&63.5&56&97.6&\textbf{9}&\textbf{76.2}&108&90.4&126&91.9&38\\
Mistique-30&84.4 &\textbf{15}&100.0&10&78.0&\textbf{21}&63.5&\textbf{38}&\textbf{94.9}&10&88.6&\textbf{72}&\textbf{88.8}&\textbf{81}&\textbf{87.6}&38\\
\hline
MCTS-20 (Ours)&\textbf{39.9}&257&82.3&\textbf{246}&36.9&391&33.4&909&\textbf{35.9}&743&23.3&1280&20.4&2266&\textbf{39.7}&2628\\
MCTS-30 (Ours)&41.1&\textbf{107}&\textbf{58.4}&307&\textbf{35.4}&\textbf{331}&\textbf{32.0}&\textbf{571}&42.2&\textbf{447}&\textbf{20.1}&\textbf{1031}&\textbf{20.2}&\textbf{1417}&40.7&\textbf{1031}\\
\hline
Greedy-20 (Ours)&28.5&44&\textbf{29.4}&55&\textbf{24.4}&44&\textbf{30.0}&60&\textbf{33.3}&48&\textbf{16.1}&99&\textbf{18.6}&135&\textbf{37.2}&228\\
Greedy-30 (Ours)&\textbf{25.1}&\textbf{32}&30.4&\textbf{40}&29.9&\textbf{32}&32.0&\textbf{40}&36.9&\textbf{32}&16.2&\textbf{72}&20.8&\textbf{90}&40.8&\textbf{152}\\
\hline
\texttt{DRD}  (Ours)&\textbf{29.7}&\textbf{18}&\textbf{27.3}&\textbf{30}&24.4&\textbf{16}&34.1&\textbf{19}&35.5&\textbf{21}&\textbf{17.0}&\textbf{55}&24.7&\textbf{50}&\textbf{37.6}&\textbf{114}\\
\texttt{DRED} (Ours) &31.6&26&35.6&36&\textbf{23.9}&20&\textbf{33.4}&26&\textbf{34.2}&25&17.7&59&\textbf{24.6}&56&37.7&149\\
\hline
\end{tabular}
}
\end{center}
\end{table*}

\subsubsection{Overall Results}
\textcolor{black}{We first compared the \textbf{compression ratio} and the required \textbf{number of validations} of various baselines on scenarios A1-A5 and B1-B3. The results, presented in Tab.~\ref{tab:tasks}, showed that our \texttt{DRD} and \texttt{DRED} achieved $1.9$ to $3.7\times$ better compression ratio than Mistique, and $1.3$ to $3.3\times$ better than Dedup. Dedup and Mistique sometimes achieved poor compression ratios (e.g., $>95\%$ for A2 and A3) with very small validation numbers. That is because of their early stop mechanism (i.e., immediately stop in case of a validation failure). We also found that MCTS requires numerous validation steps yet fails to achieve a low compression ratio due to the large search space. For Greedy-N, it is challenging to automatically tune the value of $N$. When we tune $N$ to achieve a competitive compression ratio, the validation number (a constant $|M_t|/N$) is usually worse than \texttt{DRD} and \texttt{DRED}. 
In conclusion, our \texttt{DRD} and \texttt{DRED} achieved better trade-offs between the model compression ratio and the validation times.}

\noindent
\textbf{Deduplication Overhead.} Note that most of the deduplication time is spent on accuracy validation. For example, each validation takes $85.6$ seconds for Roberta on QNLI, $74.4$ seconds for Roberta on SST2, $58.3$ seconds for Roberta on one MNLI partition, $84.7$ seconds for ViT on CIFAR100, and $84.8$ seconds for ResNet on CelebA. It highlights the importance of balancing both objectives.

\begin{figure}[ht]
\vspace{-5pt}
\centering
\includegraphics[width=0.49\textwidth]{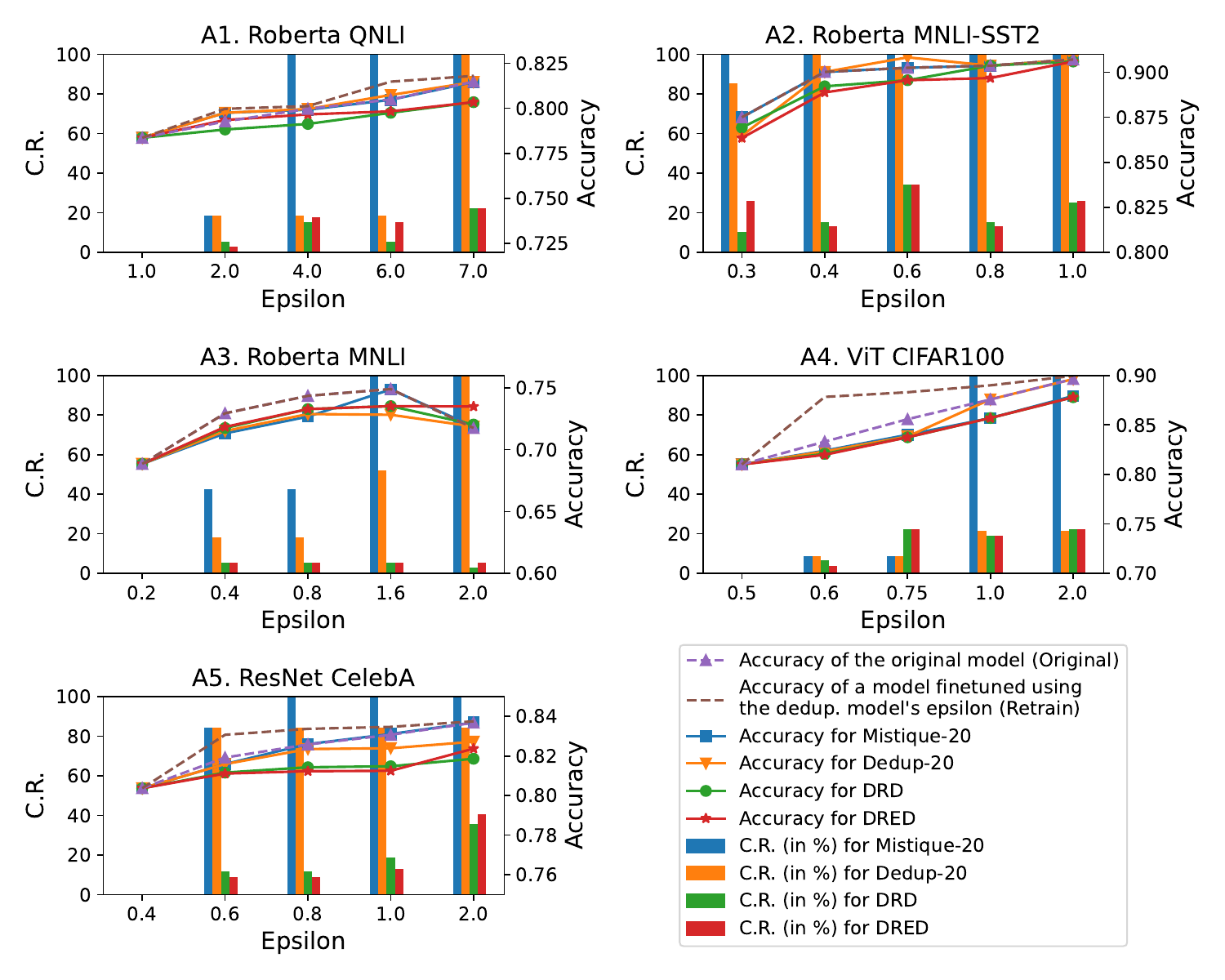}
\caption{\textcolor{black}{Compression ratio (C.R.) for individual models in A1 to A5. 
C.R. of the base model is not affected by deduplication and is thus not shown. A2's base model is the first model in A3.}}
\label{fig:effectiveness}
\end{figure}

\noindent
\textbf{Impact to Individual Models.} Fig.~\ref{fig:effectiveness} illustrated each model's \textbf{compression ratio} and \textbf{accuracy} after applying various deduplication approaches  to individual models in A1-A5 of Tab.~\ref{tab:scenarios}. 
%
The results confirmed the storage benefits achieved by our proposed \texttt{DRD} and \texttt{DRED} methods.
Considering individual models, DRED outperformed Dedup-20 and Mistique-20 by up to $35\times$.
 In addition, the Retrain baseline (i.e., full fine-tuning) achieved the best accuracy within the same privacy constraint. However, it does not perform any deduplication. Other baselines achieved similar accuracy to ours since we are using the same accuracy constraints, except that in a few cases where, Dedup and Mistique stopped early due to an accuracy validation failure, leading to high accuracy accompanied by a significantly worse compression ratio (i.e., close to $100\%$, which means zero compression) than our approaches.

\begin{figure}[ht]
\centering
\vspace{-5pt}
\includegraphics[width=0.46\textwidth]{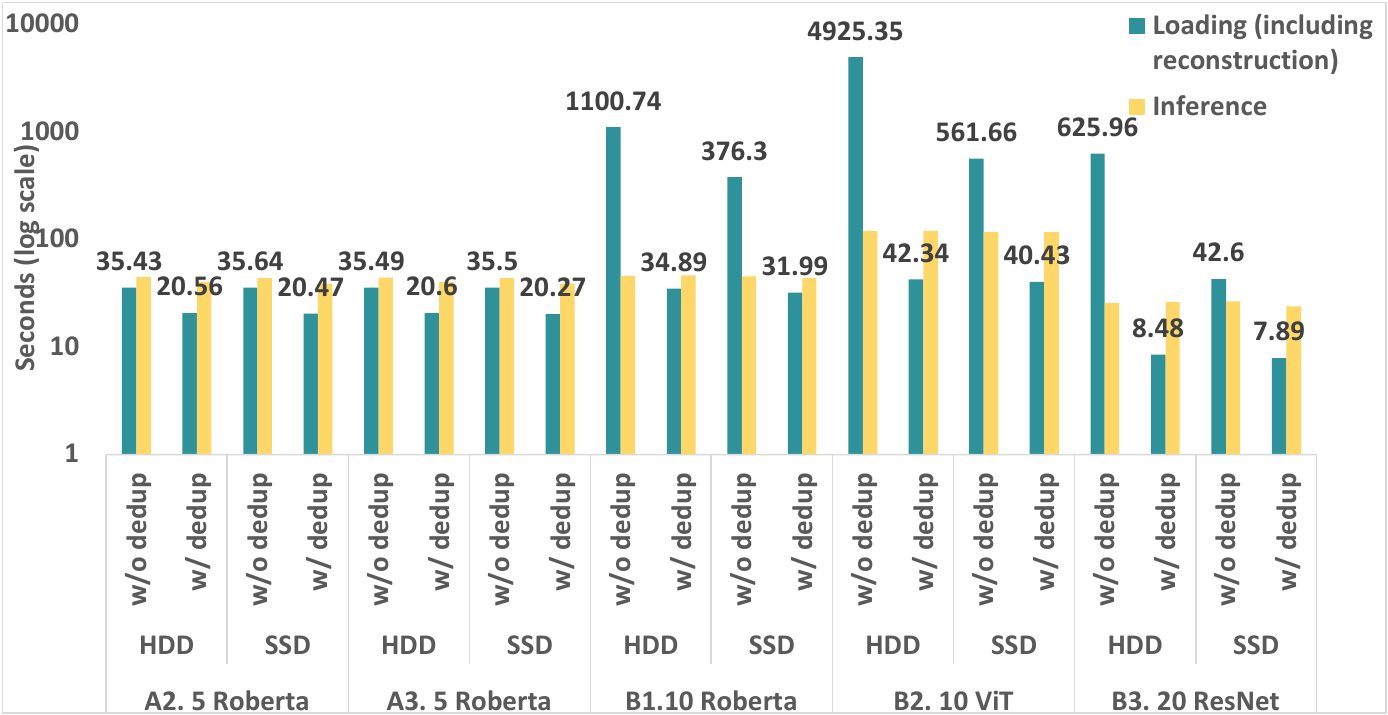}
\caption{\textcolor{black}{Latency breakdown of serving $100$ inference queries involving multiple models randomly w/o and w/ deduplication (\texttt{DRD} is used, and latency is represented in log scale)}} 
\label{fig:latency}
\end{figure}

\noindent
 \textbf{Up to $\textbf{31}\times$ model loading and inference latency} \textcolor{black}{As illustrated in Fig.~\ref{fig:latency}, we compared the latency with and without applying our proposed deduplication approach (based on \texttt{DRD}) for A2, A3, and B1-B3, in  AWS c5a.xlarge instance with EBS gp2 SSD and EBS magnetic HDD, as described in Sec.~\ref{sec:environment}. (A1, A4, and A5 are subsets of B1, B2, and B3, and their performance in this experiment is similar to A2 and A3, which are omitted due to space limits.)} For this experiment, we measured and broke down the overall latency into loading (including reconstruction) and inference latency for serving $100$ inference queries for each scenario. Each query involves a model randomly sampled from the models in the corresponding scenario. \textcolor{black}{The results showed that our approach achieved significantly better speedups in inference latency for scenarios that involve more models: It achieved $5.6\times$ and $14.1\times$ speedup of the overall latency in SSD and HDD respectively for serving $10$ Roberta models in B1, $4.3\times$ to $31.0\times$ for serving $10$ ViT models in B2, and $2.2\times$ to $18.8\times$ for serving $20$ ResNet models in B3.}
 \textcolor{black}{Even for smaller scenarios such as A2 and A3, our deduplication approach achieved around $1.3\times$ speedup for both SSD and HDD.} 
 The results demonstrated that by deduplicating the models' weights, the overall memory footprint required to serve multiple models can be significantly reduced, bringing lower I/O overheads and lower cache misses. 

\vspace{-5pt}
\subsubsection{Base Model Selection}
\label{sec:exp-base-model}
\begin{figure}[ht]
\vspace{-5pt}
\centering
\includegraphics[width=0.48\textwidth]{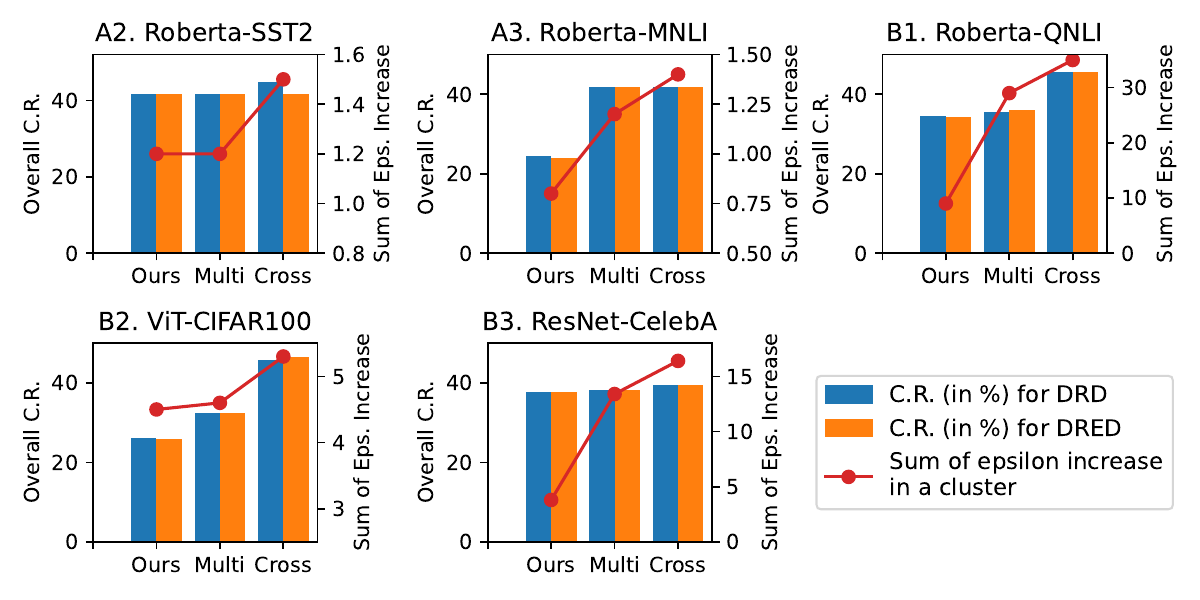}
\caption{\textcolor{black}{Comparison of Base Model Selection Strategies.}
} 
\label{fig:base-model}
\vspace{-5pt}
\end{figure}

\textcolor{black}{We compared the overall compression ratio and increased privacy costs of our base model selection algorithm (Alg.~\ref{alg:base-model-main}) to the Multi and Cross baselines on five distinct scenarios A2, A3, and B1 to B3, as shown in Fig.~\ref{fig:base-model}. To allow Multi and Cross, we relax the $\epsilon^*$ of each model to be the sum of the first three models in its scenario, while $u^*$ is kept the same.
Our Alg.~\ref{alg:base-model-main} achieved the lowest
 overall increase in privacy costs, which is $1.7$ to $4.3\times$ lower than Cross and $1.0$ to $3.5\times$ lower than Multi. Our approach also achieved $1$ to $1.7\times$ and $1$ to $1.8\times$ better compression ratios than Multi and Cross respectively. Taking A3 as an example, it achieved $1.7\times$ better compression ratio and $1.5\times$ and $1.75\times$ lower privacy costs compared to Multi and Cross.}






\begin{table*}[]
\begin{center}
\scriptsize
{\color{black}
\caption{\textcolor{black}{Comparison of Deduplication Algorithms With SVT.}}
\label{tab:tasks-svt}
\begin{tabular}{|c||c|c||c|c||c|c||c|c||c|c||c|c||c|c||c|c|}
\hline
&\multicolumn{2}{|c||}{A1}  &\multicolumn{2}{|c||}{A2} &\multicolumn{2}{|c||}{A3}
&\multicolumn{2}{|c||}{A4} 
&\multicolumn{2}{|c||}{A5}
&\multicolumn{2}{|c||}{B1}
&\multicolumn{2}{|c||}{B2}
&\multicolumn{2}{|c|}{B3}\\
&C.R.(\%)&\#Val.&C.R.(\%)&\#Val.&C.R.(\%)&\#Val.&C.R.(\%)&\#Val.&C.R.(\%)&\#Val.&C.R.(\%)&\#Val.&C.R.(\%)&\#Val.&C.R.(\%)&\#Val.\\ 
\hline
\hline
Greedy-20&28.9&44&\textbf{72.7}&22&\textbf{27.3}&44&\textbf{35.9}&56&\textbf{42.1}&47&18.0&97&26.6&127&\textbf{40.3}&206\\
Greedy-30&28.9&\textbf{32}&77.3&\textbf{20}&29.9&\textbf{32}&42.8&\textbf{38}&50.0&\textbf{32}&\textbf{16.3}&\textbf{72}&\textbf{24.3}&\textbf{90}&42.8&\textbf{148}\\
\hline
\texttt{DRD}&33.6&\textbf{24}&70.4&21&24.9&\textbf{20}&37.8&24&\textbf{38.1}&\textbf{21}&\textbf{18.0}&51&\textbf{28.8}&51&\textbf{42.3}&113 \\
\texttt{DRED}&\textbf{32.1}&28&\textbf{69.9}&\textbf{17}&\textbf{24.2}&26&\textbf{35.3}&\textbf{21}&40.4&24&19.0&\textbf{49}&29.1&\textbf{46}&43.5&\textbf{96}\\
\hline
\end{tabular}
}
\vspace{-5pt}
\end{center}
\end{table*}

\subsubsection{SVT for Private Data Validation}
\label{sec:exp-svt}
\begin{figure}[ht]
\centering
\includegraphics[width=0.48\textwidth]{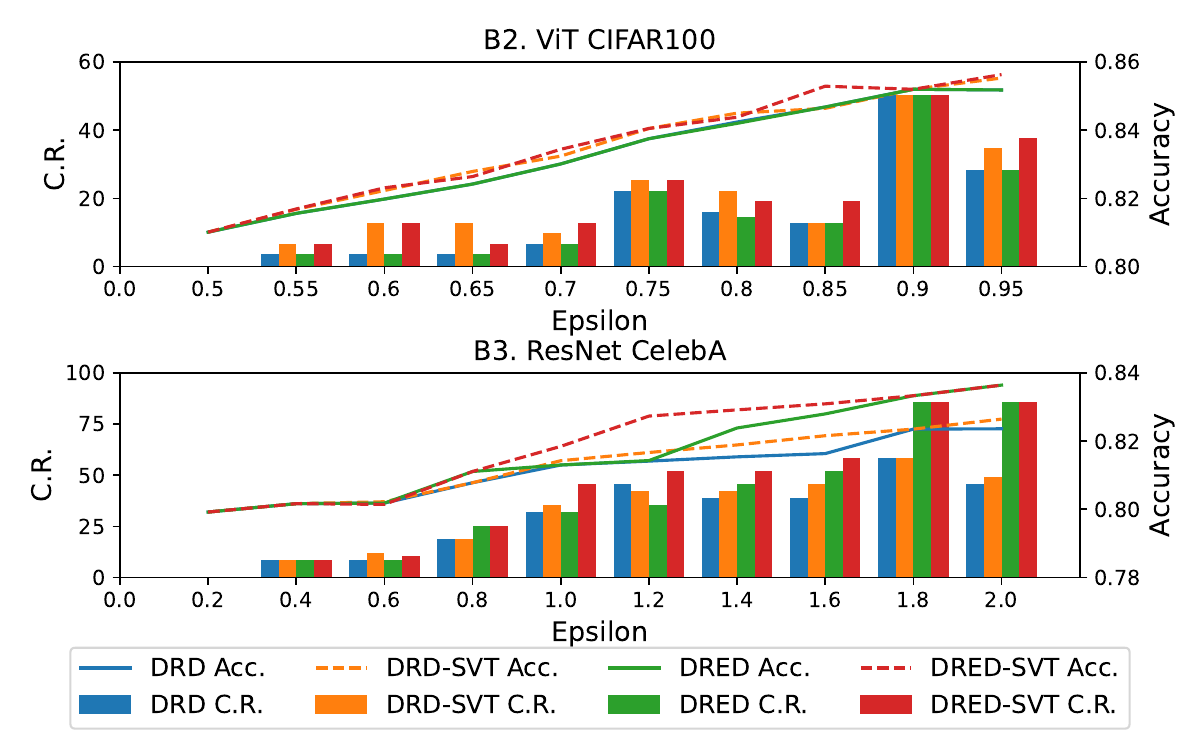}
\caption{\textcolor{black}{Deduplication with SVT-based accuracy validation. The compression ratio (C.R.) of the base models, is not not shown.}}
\label{fig:svt}
\end{figure}

We next investigated how using the Sparse Vector Technique (SVT) for model evaluation on private validation datasets affects compression ratio and accuracy.  We set the SVT cut-off to $3$ and allocated the privacy budget for SVT as the sum of the base and target model budgets, ensuring no additional privacy cost over the entire dataset. 
\textcolor{black}{We present the overall results for scenarios A1-A5, and B1-B3 
in Tab.~\ref{tab:tasks-svt}, using Greedy-N, \texttt{DRD}, and \texttt{DRED}. The comparison between Tab.~\ref{tab:tasks} and Tab.~\ref{tab:tasks-svt} showed that the number of validation steps of the Greedy-N algorithm is more constrained if SVT is used, which worsened the compression ratio, leading to a gap ranging from $0.1\%$ (on A3) to $46.9\%$ (on A5, from $30.4\%$ to $77.3\%$).
However, \texttt{DRD} and \texttt{DRED} are more robust to SVT's cutoff on the validation failures, because of their dynamic range selection without compromising compression ratio, for which the gap between the compression ratios w/o SVT and w/ SVT ranges from $-4.0\%$ (on A5) to $4.5\%$ (on B2).
}
%
%
\textcolor{black}{We also showed the  impacts to individual models on B2 and B3 scenarios in Fig.~\ref{fig:svt}. The results confirmed that SVT usually worsens the compression ratio by up to \textcolor{black}{$16.7\%$} and in certain cases it improves the compression ratio by up to \textcolor{black}{$3.2\%$} (due to randomness introduced by the noisy utility drop comparison). Observations in other scenarios are similar.}

\vspace{-10pt}

\begin{table*}[h]
\begin{center}
\color{black}
\scriptsize
\caption{\textcolor{black}{Comparison of different deduplication algorithms for the multi-user many-model scenario including $50$ models.}
}
\label{tab:realistic-multiuser}
\begin{tabular}{|c||c|c||c|c||c|c||c|c||c|c|}
\hline
&Dedup-20&Dedup-30&Mistique-20&Mistique-30&MCTS-20 (Ours)&MCTS-30 (Ours)&Greedy-20 (Ours)&Greedy-30 (Ours)&\texttt{DRD} (Ours)&\texttt{DRED} (Ours)\\ 
\hline
\hline
C.R. (\%) &$57.1$&$46.7$&$87.1$&$88.6$&$32.2$&$28.3$&$\textbf{21.2}$&$23.4$&$23.7$&$24.9$\\
\#Val &$327$&$257$&$308$&$222$&$6811$&$4117$&$561$&$386$&$264$&$320$\\
\hline
\end{tabular}
\end{center}
\end{table*}

\begin{table*}[h]
\centering
\color{black}
\scriptsize
\caption{\label{tab:realistic-single-user} {\textcolor{black}{Comparison of different deduplication algorithm for the single-user heterogeneous-model scenario. 
}}}
\begin{tabular}{|c||c|c||c|c||c|c||c|c||c|c||c|c||c|c|} \hline
\multirow{3}{4em}{Datasets}
&\multicolumn{2}{|c||}{Roberta-SST2 ($\epsilon=5$)}&\multicolumn{2}{|c||}{Roberta-IMDB ($\epsilon=5$)}&\multicolumn{2}{c||}{Roberta-QNLI ($\epsilon=5$)}&\multicolumn{2}{c||}{ViT-GTSRB ($\epsilon=2$)}&\multicolumn{2}{c||}{ViT-CelebA ($\epsilon=2$)}&\multicolumn{2}{c||}{ViT-SVHN ($\epsilon=2$)}&\multicolumn{2}{c|}{Overall}\\
&C.R.(\%)&\#Val.&C.R.(\%)&\#Val.&C.R.(\%)&\#Val.&C.R.(\%)&\#Val.&C.R.(\%)&\#Val.&C.R.(\%)&\#Val.&C.R.(\%)&\#Val. \\
\hline \hline
~\texttt{Greedy-1}&63.4&210&69.5&211&68.6&210&54.2&578&15.9&578&54.8&578&58.8&2365 \\ \hline
~\texttt{Greedy-20}&90.3&11&90.8&11&90.3&11&73.7&29&23.1&29&77.6&29&73.8&120 \\ \hline
~\texttt{Greedy-30}&99.8&7&100.0&8&85.6&7&77.6&20&25.1&20&77.6&20&76.0&82 \\ \hline
~\texttt{DRD}&87.5&5&88.0&5&84.7&8&77.6&11&21.4&8&79.3&13&73.6&50 \\ \hline
~\texttt{DRED}&87.5&8&88.0&8&82.8&18&78.2&18&23.5&13&78.2&18&73.8&83 \\ \hline
\end{tabular}
\end{table*}

\vspace{-5pt}
{\color{black}
\subsection{Two Realistic Scenarios}
\subsubsection{Multi-User Many-Model Scenario}
The overall deduplication effectiveness w/ public validation datasets (w/o SVT-based validation) is shown in Tab.~\ref{tab:realistic-multiuser}. Greedy-20 achieved the best compression ratio of $21.2\%$. However, it requires $561$ validations, representing a long latency of more than $10$ hours; Mistique-30 achieved the lowest number of validations, $222$, but its compression ratio is not ideal, $88.6\%$. We found our \texttt{DRD} method achieves the best trade-off between compression ratio and validation overheads, with a near-optimal compression ratio of $24.5\%$ and a near-optimal validation number of $264$. The results on private validation dataset (w/ SVT-based validation) is similar, the \texttt{DRD} and \texttt{DRED} achieved similar effective compression ratios ($32.0\%$ and $32.4\%$) with Greedy-20 and Greedy-30 ($31.4\%$ for both), while the former's validation numbers ($256$ and $234$) are significantly lower than the latter ($496$ and $362$).

When serving all $50$ models concurrently on a c5a.2xlarge instance, the end-to-end latency (including total model loading and inferences) for all $100$ inference queries (each query randomly selects one model for inference) are shown in the left part of Fig.~\ref{fig:realistic-scenario-latency}. Deduplication achieved a $28\times$ speedup when HDD is used, and $11\times$ speedup when SSD is used.

\subsubsection{Single-User Heterogeneous-Model Scenario}
Tab.~\ref{tab:realistic-single-user} demonstrates that our deduplication algorithms (w/o SVT-based validation), such as \texttt{DRD} is able to achieve benefits even for multiple heterogeneous models. \texttt{DRD} outperformed all baselines in minimizing the total required number of validations ($50$). It also achieved the second-best compression ratio ($73.6\%$). While Greedy-1 achieved the best compression ratio ($58.8\%$), it is at the cost of $2,365$ validations (i.e., more than $50$ hours). On the contrary, \texttt{DRD} only requires $50$ validations, which is a $47\times$ speedup. 

When serving all $6$ models concurrently on a c5a.1xlarge instance, the end-to-end latency for all $100$ random inference queries (each query randomly selects one model for inference) is illustrated in the right part of Fig.~\ref{fig:realistic-scenario-latency}. Deduplicated models achieved a $43\times$ speedup of the overall inference time on HDD storage and a $14\times$ speedup, compared to non-deduplicated models.

 \begin{figure}[ht]
\centering
\includegraphics[width=0.46\textwidth]{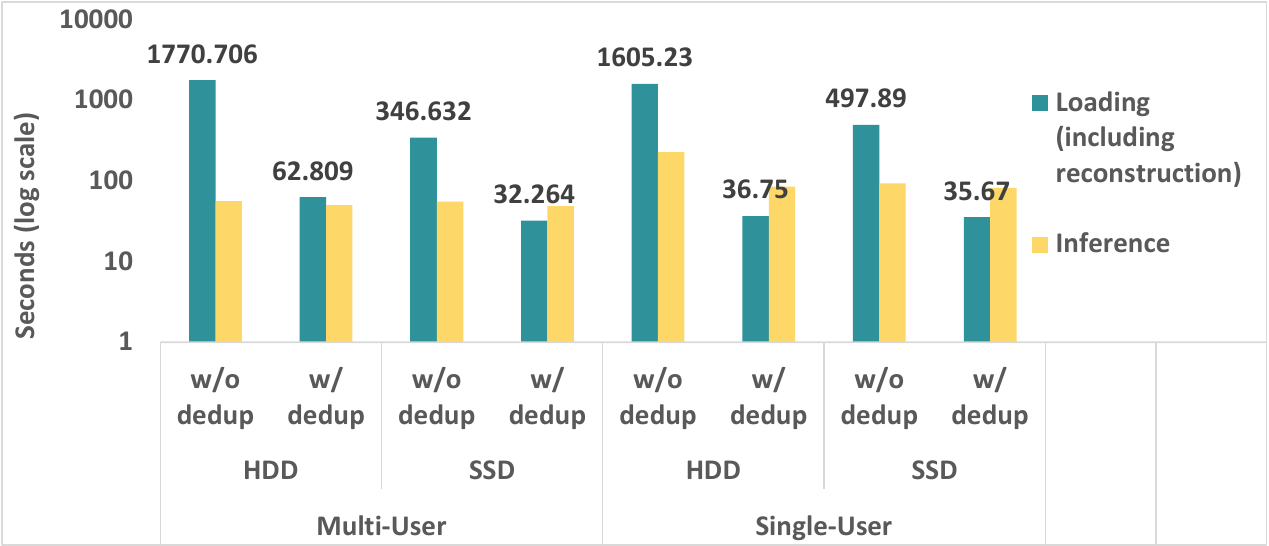}
\caption{\textcolor{black}{Latency breakdown of serving $100$ random inference queries for the realistic scenarios w/o and w/ deduplication (\texttt{DRD} is used, and latency is represented in log scale)}}
\label{fig:realistic-scenario-latency}
\end{figure}

}




\vspace{-15pt}
\subsection{Ablation Studies}
\label{sec:exp-ablation}
\subsubsection{Impact of Deduplication Hyper-Parameters} We compared the impact of block size, saliency measurement, saliency aggregation method, and distance measure. All experiments in this section used the ~\texttt{DRED} algorithm.
We first measured how varying block sizes affect compression ratio, validation steps, and accuracy drop using scenario A4, which consists of five ViT models. We observed that smaller blocks usually resulted in better compression ratios. However, smaller block sizes also lead to more blocks per model, thereby increasing the number of validation steps. In addition, when the block size decreases to a point, the compression ratio will not improve anymore. This is because salient weights exhibit locality and are distributed in multiple small clusters, as verified by Lee et al.~\cite{lee2020fast}. When the block size is smaller than the size of these clusters, the compression ratio does not improve.

We further compared the effectiveness of different saliency measures, the saliency aggregation methods, and block similarity measurements, using the deduplication of the A1 scenario as an example. The results in Tab.~\ref{tab:ablation-study} showed that our proposed gradient magnitude measurement outperformed other measurements regarding compression ratio. We also find that profiling of the weight magnitude is the most efficient, taking $113$ seconds, while calculating the Wanda score, Fisher information, and gradients using the entire validation dataset takes $145$ seconds, $6,576$ seconds, and $1,060$ seconds, respectively. For saliency aggregation,
we found using the l2-norm of weight gradients as the block saliency outperformed l1-norm, l-infinite, and third quartile of weight gradients, as illustrated in Tab.~\ref{tab:ablation-study}. 
For measuring the similarity of two blocks, as shown in Tab.~\ref{tab:ablation-study},  l2 (Eclidean) distance outperformed other metrics.

\begin{table}[h]
\vspace{-5pt}
\centering
\scriptsize
\caption{\label{tab:ablation-study} {Ablation Study of Deduplication Hyper-Parameters}}
\begin{tabular}{|c|c|c|c|c|c|c|c|} \hline
&Values&C.R.(\%)&\#Val.&$\max{(\Delta u)}$\\ \hline \hline
\multirow{5}{1.5cm}{A4. block size (\#floating points)} 
&4,194,304&53.6&9&0.016 \\ 
&2,097,152&41.1&16&0.017 \\ 
&1,048,576&33.4&26&0.018 \\ 
&524,288&32.5&29&0.020 \\ 
&262,144&32.12&29&0.020 \\ \hline
\multirow{ 4}{1.5cm}{A1. saliency measurement}&Weight Magnitude&63.8&49&0.015 \\ 
&Wanda&47.3&30&0.015 \\ 
&Fisher Information&45.9&24&0.015 \\ 
&Gradient Magnitude&32.8&31&0.019 \\ \hline
\multirow{ 4}{1.5cm}{A1. saliency aggregation method}&l1-norm&27.6&26&0.014 \\ 
&l2-norm&26.1&31&0.013 \\ 
&l-infinite&30.0&35&0.015 \\ 
&3rd-quartile&27.1&32&0.015 \\ \hline
\multirow{ 3}{1.5cm}{A1. pairwise distance measurement}&l1-distance&25.9&29&0.014 \\
&l2-distance&40.2&31&0.012 \\ 
&cosine&92.1&63&0.014 \\ \hline
\end{tabular}
\vspace{-5pt}
\end{table}

\vspace{-5pt}
\subsubsection{Impact of Accuracy Threshold}
We used various accuracy drop thresholds to deduplicate models in A1, A4, and A5 scenarios, and recorded the compression ratios shown as percentages in Tab.~\ref{tab:ablation-threshold}.  We found that generally, larger accuracy drop thresholds yield better compression ratios. In addition, a negative accuracy drop threshold requires the resulting model to have a higher accuracy than the original one. When the threshold was set to $-0.5\%$, no blocks could be deduplicated in A4 and A5. However, for A1, a few blocks could be deduplicated to improve the accuracy by 0.5\%, which indicates that although we focus on constraining the utility drop, deduplication may improve utility in certain cases. 


\begin{table}[h]
\vspace{-5pt}
\centering
\scriptsize
\caption{\label{tab:ablation-threshold} {Compression Ratios for Different Accuracy Thresholds}}
\begin{tabular}{|c|c|c|c|c|c|c|c|} \hline
Thresholds(\%)&-0.5&0.0&0.5&1.0&1.5&2.0\\ \hline \hline
A1. Roberta&83.7&58.0&23.4&21.9&21.1&20.8 \\ \hline
A4. ViT&N/A&84.4&54.1&41.5&35.9&31.0 \\ \hline
A5. ResNet&N/A&80.2&44.8&37.8&35.1&31.2 \\ \hline
\end{tabular}
\end{table}

\subsubsection{Impacts of SVT hyper-parameter tuning.} 
We also investigated the impacts of cut-off $c$ in SVT-based validation. 
(Our configuration of privacy budget for SVT-based validation is consistent with Sec.~\ref{sec:exp-svt}.) 
 Taking the deduplication in the A1 scenario as an example, we varied the value of $c$ from $2$ to $7$, and recorded the compression ratio, the maximum accuracy drop, and the number of validations in Tab.~\ref{tab:cut-off}. The results showed that when $c$ reaches a point (e.g., $3$ in the example), increasing it will not improve the compression ratio, although a larger cut-off allows for more validation failures before it terminates the deduplication process. That's because increasing $c$ will raise the noise level added to the query result, leading to less accurate comparisons. 

\begin{table}[h]
\vspace{-5pt}
\centering
\scriptsize
\caption{\label{tab:cut-off} {Impacts of SVT's Cut-off for the A4 cluster}}
\begin{tabular}{|c|c|c|c|c|c|c|c|} \hline
Cut-off $c$&2&3&4&5&6&7\\ \hline \hline
C.R.(\%)&33.6&31.7&31.7&31.7&31.7&31.7 \\ \hline
$\max{(\Delta u)}$(\%)&0.66&1.14&1.14&1.14&1.14&1.14 \\ \hline
Num. of val.&18&23&26&28&28&28 \\ \hline
\end{tabular}
\vspace{-5pt}
\end{table}

\begin{figure}[ht]
\centering
\vspace{-5pt}
\includegraphics[width=0.48\textwidth]{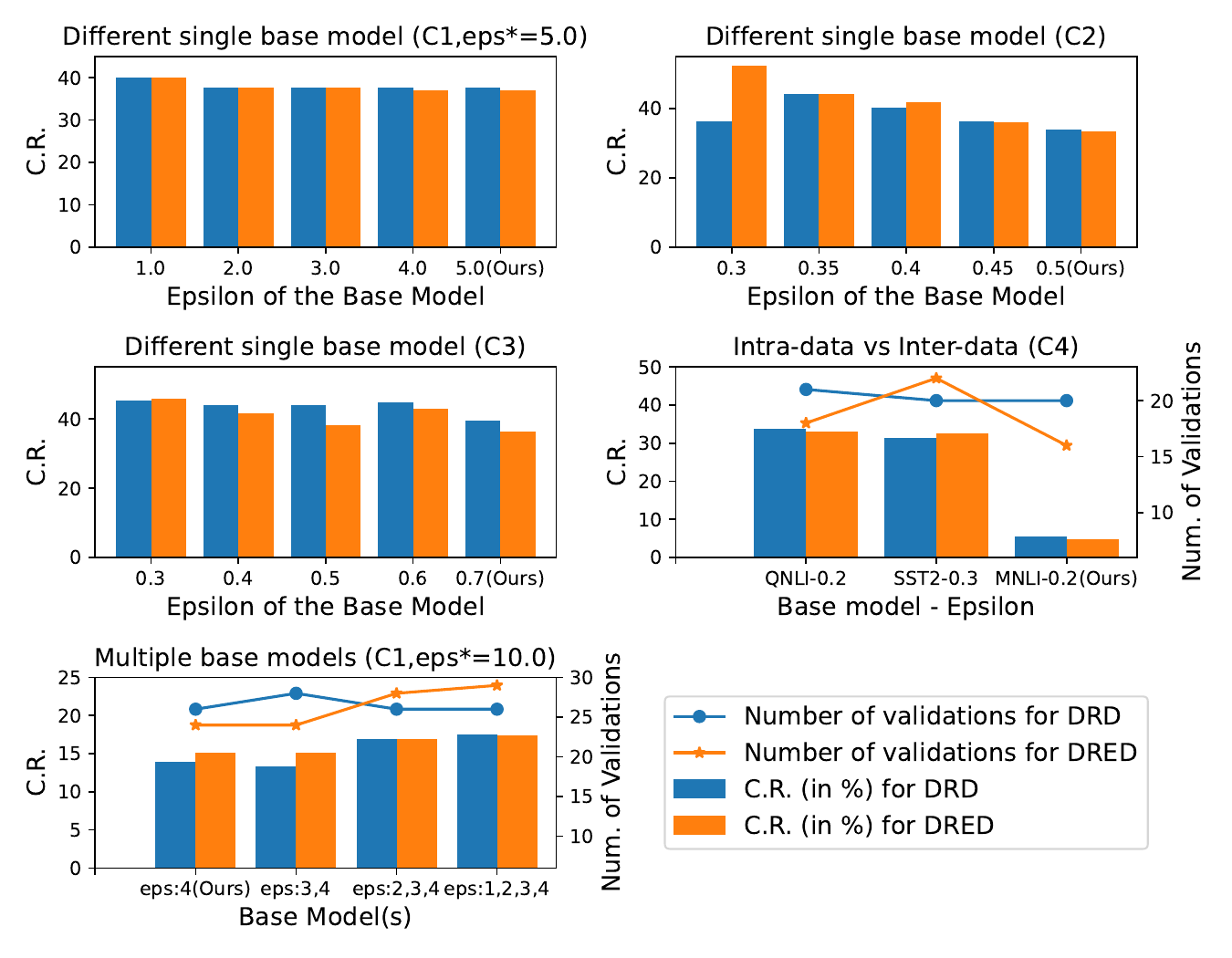}
\caption{\textcolor{black}{Ablation study for base model selections.}}
\vspace{-5pt}
\label{fig:bms_ablation}
\end{figure}

\subsubsection{Impacts of base model selection}

We further investigated how the selection of different base models in each group will affect the compression ratio and number of validations (i.e., deduplication efficiency) in three different situations. \textit{(1) Base model selection for models trained on the same dataset with different privacy budgets} evaluated on scenarios C1-C3. For each scenario, we compared the overall compression ratio achieved in deduplicating all models from this scenario using different base models with their $\epsilon$ specified in the $x$-axis. Among these base models, only the model marked with "Ours" was selected by our algorithm. Other base models were trained for comparison.
As presented in Fig.~\ref{fig:base-model}, when the base model's epsilon increased, the compression ratio improved (i.e., decreased), which is expected since the similarity to the target models (i.e., similarity of privacy budgets) improved.
\textit{(2) Base model selection for models trained on different datasets.} For scenario C4 with multiple models trained on disjoint MNLI partitions, we found using a base model trained on a similar dataset (e.g., MNLI) outperformed other candidate base models.
\textit{(3) Using multiple base models for a group.} %
In C1, if we change the privacy constraint into $10.0$, as shown in the last plot, using a model with $\epsilon=4$ (selected by our algorithm) as the base model achieves a better compression than using three or four base models, which showed that the quality of base models is more important than quantity.

\vspace{-5pt}
{\color{black}
\subsection{Summary of Findings}
Overall, our proposed work brings $17\%$ to $38\%$ C.R. (compression ratio) for eight representative scenarios, leading to a $31\times$ model serving latency speedup. In addition, in a realistic scenario that involves $50$ models, our work achieved $23.7\%$ C.R., resulting in $28\times$ serving speedup. Even in a challenging scenario that runs multiple heterogeneous models required by a social robot in a resource-constrained environment, our approach achieved an impressive $43\times$ inference latency speedup brought by a C.R. of $74\%$. 

In addition, our dynamic deduplication algorithms \texttt{DRD} and \texttt{DRED} have improved the compression ratio by up to $3.3\times$ compared to the best of existing deduplication algorithms(, which do not consider privacy constraints at all). Moreover, our base model selection strategy has reduced privacy costs by up to $3\times$ compared to alternative base model selection designs. Our SVT-based accuracy validation strategy is also demonstrated as effective in minimizing the privacy loss on private validation datasets, with negligible impacts on compression ratio if coupled with \texttt{DRD} and \texttt{DRED}.

Our other observations include:

\noindent
\textbf{Deduplication Algorithm.} Our proposed \texttt{DRD} and \texttt{DRED} algorithms provide the optimal trade-off between compression ratio and deduplication overhead, compared to all other baselines.  Greedy-1 can achieve the optimal compression ratio at the cost of inhibiting deduplication latency.

\noindent
\textbf{Block Size Selection.} A smaller block size usually leads to a better compression ratio with higher deduplication overhead. However, a very small block size not only slows down the deduplication, but also overlooks the synergy among adjacent blocks, thus not benefiting the compression ratio. 

\noindent
\textbf{Saliency Measure.} Weight magnitude is easy to measure but its accuracy is sub-optimal. Wanda~\cite{sun2024simple} based on weights and activation is only applicable to linear layers.
Fisher information~\cite{liu2021group} suffers from the value vanishing problem caused by the square of small gradient values. Our proposed gradient magnitude measure is more effective in our target problems than the above approaches.

\noindent
\textbf{Base Model Selection} Compared to our proposed base model selection strategy, baselines such as Multi and Cross do not improve the compression ratios. However, these baselines significantly increase the composed model-specific privacy costs because they introduce more low-quality base models. 

\noindent
\textbf{SVT Cut-off} A smaller cutoff incurs smaller noise. However, it limits the maximum number of failure during deduplication. A default choice of $3.0$ works well in most of our experiments.
}


%% file: discussion.tex
\vspace{-5pt}
{\color{black}{

\section{Further Discussion}
In this section, we discuss the scalability of our proposed approach and the dynamic addition/removal of models. 

\noindent
\textbf{Scalability of the Deduplication Process.} First, each pair of base and target models can be deduplicated in parallel with other pairs. 
Second, the deduplication of a target model ($M_j$) using a  base model ($M_i$) can be split into two phases: (1) running the deduplication algorithm, where the peak memory is $mem\_dedup = size(M_i)+block\_size$, if the base model $M_i$ is cached in memory; and (2) validating the accuracy, with peak memory being $mem\_valid = size(M_j)+size(feature\_maps)+size(input)$, which mainly sums up the size of the target model, the intermediate feature maps created for inference, and the validation data. Therefore, we have $peak\_{mem} = \max(mem$ $\_dedup, mem\_valid)$.

\noindent
\textbf{Model Addition/Removal.}
Our algorithm can be extended to handle dynamic model addition and removal. A batch of newly arrived models will be dispatched to existing or new clusters based on metadata similarity. Then, it will apply Alg.~\ref{alg:base-model-step} to each new model.
For model removal, the process depends on whether the model is a base model. Non-base models can be safely removed without affecting the system. However, if a base model is deemed no longer useful, two strategies can be employed. The first strategy is to retain its blocks that are in use by other models, minimizing disruption. The second strategy involves archiving the original models in cold storage and re-run the deduplication process for affected models, using an alternative base model.
}}

%% file: relatedworks.tex
\vspace{-5pt}
\section{Other Related Works}
\label{sec:survey}

\textcolor{black}{
Both SmartLite ~\cite{lin2023smartlite}
and Nexus~\cite{shen2019nexus} propose to reuse these shared layers of multiple models fine-tuned from the same pre-trained model. However, they did not consider deduplicating similar but non-identical tensor blocks, which will significantly improve the compression ratio in a broad class of scenarios, e.g., full-parameter fine-tuning, heterogeneous models, models with multiple DP versions, etc.}
 Weight virtualization~\cite{lee2020fast} merges pages across multiple models into a single page.  \textcolor{black}{Model merging~\cite{yang2024model} builds a universal model that excels at multiple tasks by merging the weights of multiple models.}  However, none of these works have considered privacy.
 Deduplication of relational data in RDBMS, also known as record linkage, identifies duplicate items through entity matching~\cite{elmagarmid2006duplicate,mests2018distributed,thirumuruganathan2021deep}, using various blocking techniques to avoid the pair-wise comparison for dissimilar items~\cite{bilenko2006adaptive,  ananthakrishna2002eliminating, hernandez1995merge, borthwick2020scalable,chu2016distributed,kolb2012load, kolb2012dedoop}.}
In addition, various techniques leveraged similarity functions to filter out pairs that have similarity scores below a threshold~\cite{xiao2008ed} or used LSH to convert similarity join to an equi-join problem~\cite{yu2016generic}. \textcolor{black}{However, these works did not consider how the deduplication will affect the accuracy and privacy of downstream ML applications.} 

%% file: conclusion.tex
\section{Conclusions}
This work proposed a privacy-centric redesign of model deduplication techniques to compress multiple models and alleviate the inference and storage costs of multi-tenant and multi-tasking model serving platforms in resource-constrained environments. We identified and formalized a novel accuracy- and privacy-aware model deduplication problem. To address the problem, we proposed novel techniques including a greedy base model selection strategy to optimize the privacy costs and novel deduplication strategies (e.g., DRD/DRED) that deduplicate a dynamic number of blocks each time based on an ordering of blocks by saliency. 
We also leverage SVT to reduce privacy costs of validating accuracy using private datasets. We conducted detailed evaluations on diverse and representative clusters of models and obtained promising results. 


%% file: main.bbl

\begin{thebibliography}{89}


\ifx \showCODEN    \undefined \def \showCODEN     #1{\unskip}     \fi
\ifx \showDOI      \undefined \def \showDOI       #1{#1}\fi
\ifx \showISBNx    \undefined \def \showISBNx     #1{\unskip}     \fi
\ifx \showISBNxiii \undefined \def \showISBNxiii  #1{\unskip}     \fi
\ifx \showISSN     \undefined \def \showISSN      #1{\unskip}     \fi
\ifx \showLCCN     \undefined \def \showLCCN      #1{\unskip}     \fi
\ifx \shownote     \undefined \def \shownote      #1{#1}          \fi
\ifx \showarticletitle \undefined \def \showarticletitle #1{#1}   \fi
\ifx \showURL      \undefined \def \showURL       {\relax}        \fi
\providecommand\bibfield[2]{#2}
\providecommand\bibinfo[2]{#2}
\providecommand\natexlab[1]{#1}
\providecommand\showeprint[2][]{arXiv:#2}

\bibitem[Abadi et~al\mbox{.}(2016)]%
        {abadi2016deep}
\bibfield{author}{\bibinfo{person}{Martin Abadi}, \bibinfo{person}{Andy Chu}, \bibinfo{person}{Ian Goodfellow}, \bibinfo{person}{H~Brendan McMahan}, \bibinfo{person}{Ilya Mironov}, \bibinfo{person}{Kunal Talwar}, {and} \bibinfo{person}{Li Zhang}.} \bibinfo{year}{2016}\natexlab{}.
\newblock \showarticletitle{Deep learning with differential privacy}. In \bibinfo{booktitle}{\emph{CCS}}. \bibinfo{pages}{308--318}.
\newblock


\bibitem[Ananthakrishna et~al\mbox{.}(2002)]%
        {ananthakrishna2002eliminating}
\bibfield{author}{\bibinfo{person}{Rohit Ananthakrishna}, \bibinfo{person}{Surajit Chaudhuri}, {and} \bibinfo{person}{Venkatesh Ganti}.} \bibinfo{year}{2002}\natexlab{}.
\newblock \showarticletitle{Eliminating fuzzy duplicates in data warehouses}. In \bibinfo{booktitle}{\emph{VLDB'02: Proceedings of the 28th International Conference on Very Large Databases}}. Elsevier, \bibinfo{pages}{586--597}.
\newblock


\bibitem[Benitez and Malin(2010)]%
        {benitez2010evaluating}
\bibfield{author}{\bibinfo{person}{Kathleen Benitez} {and} \bibinfo{person}{Bradley Malin}.} \bibinfo{year}{2010}\natexlab{}.
\newblock \showarticletitle{Evaluating re-identification risks with respect to the HIPAA privacy rule}.
\newblock \bibinfo{journal}{\emph{Journal of the American Medical Informatics Association}} \bibinfo{volume}{17}, \bibinfo{number}{2} (\bibinfo{year}{2010}), \bibinfo{pages}{169--177}.
\newblock


\bibitem[Bernau et~al\mbox{.}(2021)]%
        {bernau2021quantifying}
\bibfield{author}{\bibinfo{person}{Daniel Bernau}, \bibinfo{person}{G{\"u}nther Eibl}, \bibinfo{person}{Philip~W Grassal}, \bibinfo{person}{Hannah Keller}, {and} \bibinfo{person}{Florian Kerschbaum}.} \bibinfo{year}{2021}\natexlab{}.
\newblock \showarticletitle{Quantifying Identifiability to Choose and Audit ǫ in Differentially Private Deep Learning}. In \bibinfo{booktitle}{\emph{Proceedings of the Conference on Very Large Databases}}.
\newblock


\bibitem[Biderman et~al\mbox{.}(2024)]%
        {biderman2024lora}
\bibfield{author}{\bibinfo{person}{Dan Biderman}, \bibinfo{person}{Jose~Gonzalez Ortiz}, \bibinfo{person}{Jacob Portes}, \bibinfo{person}{Mansheej Paul}, \bibinfo{person}{Philip Greengard}, \bibinfo{person}{Connor Jennings}, \bibinfo{person}{Daniel King}, \bibinfo{person}{Sam Havens}, \bibinfo{person}{Vitaliy Chiley}, \bibinfo{person}{Jonathan Frankle}, {et~al\mbox{.}}} \bibinfo{year}{2024}\natexlab{}.
\newblock \showarticletitle{Lora learns less and forgets less}.
\newblock \bibinfo{journal}{\emph{arXiv preprint arXiv:2405.09673}} (\bibinfo{year}{2024}).
\newblock


\bibitem[Bilenko et~al\mbox{.}(2006)]%
        {bilenko2006adaptive}
\bibfield{author}{\bibinfo{person}{Mikhail Bilenko}, \bibinfo{person}{Beena Kamath}, {and} \bibinfo{person}{Raymond~J Mooney}.} \bibinfo{year}{2006}\natexlab{}.
\newblock \showarticletitle{Adaptive blocking: Learning to scale up record linkage}. In \bibinfo{booktitle}{\emph{Sixth International Conference on Data Mining (ICDM'06)}}. IEEE, \bibinfo{pages}{87--96}.
\newblock


\bibitem[Blalock et~al\mbox{.}(2020)]%
        {blalock2020state}
\bibfield{author}{\bibinfo{person}{Davis Blalock}, \bibinfo{person}{Jose~Javier Gonzalez~Ortiz}, \bibinfo{person}{Jonathan Frankle}, {and} \bibinfo{person}{John Guttag}.} \bibinfo{year}{2020}\natexlab{}.
\newblock \showarticletitle{What is the state of neural network pruning?}
\newblock \bibinfo{journal}{\emph{Proceedings of machine learning and systems}}  \bibinfo{volume}{2} (\bibinfo{year}{2020}), \bibinfo{pages}{129--146}.
\newblock


\bibitem[Borthwick et~al\mbox{.}(2020)]%
        {borthwick2020scalable}
\bibfield{author}{\bibinfo{person}{Andrew Borthwick}, \bibinfo{person}{Stephen Ash}, \bibinfo{person}{Bin Pang}, \bibinfo{person}{Shehzad Qureshi}, {and} \bibinfo{person}{Timothy Jones}.} \bibinfo{year}{2020}\natexlab{}.
\newblock \showarticletitle{Scalable Blocking for Very Large Databases}. In \bibinfo{booktitle}{\emph{Joint European Conference on Machine Learning and Knowledge Discovery in Databases}}. Springer, \bibinfo{pages}{303--319}.
\newblock


\bibitem[Broder(1997)]%
        {broder1997resemblance}
\bibfield{author}{\bibinfo{person}{Andrei~Z Broder}.} \bibinfo{year}{1997}\natexlab{}.
\newblock \showarticletitle{On the resemblance and containment of documents}. In \bibinfo{booktitle}{\emph{Proceedings. Compression and Complexity of SEQUENCES 1997 (Cat. No. 97TB100171)}}. IEEE, \bibinfo{pages}{21--29}.
\newblock


\bibitem[Bu et~al\mbox{.}({[n.\,d.]})]%
        {budifferentially}
\bibfield{author}{\bibinfo{person}{Zhiqi Bu}, \bibinfo{person}{Yu-Xiang Wang}, \bibinfo{person}{Sheng Zha}, {and} \bibinfo{person}{George Karypis}.} \bibinfo{year}{[n.\,d.]}\natexlab{}.
\newblock \showarticletitle{Differentially Private Bias-Term Fine-tuning of Foundation Models}. In \bibinfo{booktitle}{\emph{Forty-first International Conference on Machine Learning}}.
\newblock


\bibitem[Bu et~al\mbox{.}(2022)]%
        {bu2022differentially}
\bibfield{author}{\bibinfo{person}{Zhiqi Bu}, \bibinfo{person}{Yu-Xiang Wang}, \bibinfo{person}{Sheng Zha}, {and} \bibinfo{person}{George Karypis}.} \bibinfo{year}{2022}\natexlab{}.
\newblock \showarticletitle{Differentially Private Bias-Term Fine-tuning of Foundation Models}. In \bibinfo{booktitle}{\emph{Workshop on Trustworthy and Socially Responsible Machine Learning, NeurIPS 2022}}.
\newblock


\bibitem[Cattrysse et~al\mbox{.}(1994)]%
        {cattrysse1994set}
\bibfield{author}{\bibinfo{person}{Dirk~G Cattrysse}, \bibinfo{person}{Marc Salomon}, {and} \bibinfo{person}{Luk~N Van~Wassenhove}.} \bibinfo{year}{1994}\natexlab{}.
\newblock \showarticletitle{A set partitioning heuristic for the generalized assignment problem}.
\newblock \bibinfo{journal}{\emph{European Journal of Operational Research}} \bibinfo{volume}{72}, \bibinfo{number}{1} (\bibinfo{year}{1994}), \bibinfo{pages}{167--174}.
\newblock


\bibitem[Chaslot(2010)]%
        {chaslot2010monte}
\bibfield{author}{\bibinfo{person}{Guillaume Maurice Jean-Bernard~Chaslot Chaslot}.} \bibinfo{year}{2010}\natexlab{}.
\newblock \showarticletitle{Monte-carlo tree search}.
\newblock  (\bibinfo{year}{2010}).
\newblock


\bibitem[Chen et~al\mbox{.}(2019)]%
        {chen2019towards}
\bibfield{author}{\bibinfo{person}{Lingjiao Chen}, \bibinfo{person}{Paraschos Koutris}, {and} \bibinfo{person}{Arun Kumar}.} \bibinfo{year}{2019}\natexlab{}.
\newblock \showarticletitle{Towards model-based pricing for machine learning in a data marketplace}. In \bibinfo{booktitle}{\emph{Proceedings of the 2019 international conference on management of data}}. \bibinfo{pages}{1535--1552}.
\newblock


\bibitem[Chu et~al\mbox{.}(2016)]%
        {chu2016distributed}
\bibfield{author}{\bibinfo{person}{Xu Chu}, \bibinfo{person}{Ihab~F Ilyas}, {and} \bibinfo{person}{Paraschos Koutris}.} \bibinfo{year}{2016}\natexlab{}.
\newblock \showarticletitle{Distributed data deduplication}.
\newblock \bibinfo{journal}{\emph{Proceedings of the VLDB Endowment}} \bibinfo{volume}{9}, \bibinfo{number}{11} (\bibinfo{year}{2016}), \bibinfo{pages}{864--875}.
\newblock


\bibitem[Cloud({[n.\,d.]})]%
        {google-analysis-rules}
\bibfield{author}{\bibinfo{person}{Google Cloud}.} \bibinfo{year}{[n.\,d.]}\natexlab{}.
\newblock \bibinfo{title}{Restrict data access using analysis rules}.
\newblock
\newblock
\urldef\tempurl%
\url{https://cloud.google.com/bigquery/docs/analysis-rules}
\showURL{%
Retrieved Jan 17, 2024 from \tempurl}


\bibitem[Cohen et~al\mbox{.}(2006)]%
        {cohen2006efficient}
\bibfield{author}{\bibinfo{person}{Reuven Cohen}, \bibinfo{person}{Liran Katzir}, {and} \bibinfo{person}{Danny Raz}.} \bibinfo{year}{2006}\natexlab{}.
\newblock \showarticletitle{An efficient approximation for the generalized assignment problem}.
\newblock \bibinfo{journal}{\emph{Inform. Process. Lett.}} \bibinfo{volume}{100}, \bibinfo{number}{4} (\bibinfo{year}{2006}), \bibinfo{pages}{162--166}.
\newblock


\bibitem[Crankshaw(2019)]%
        {crankshaw2019design}
\bibfield{author}{\bibinfo{person}{Daniel Crankshaw}.} \bibinfo{year}{2019}\natexlab{}.
\newblock \emph{\bibinfo{title}{The Design and Implementation of Low-Latency Prediction Serving Systems}}.
\newblock \bibinfo{thesistype}{Ph.\,D. Dissertation}. \bibinfo{school}{UC Berkeley}.
\newblock


\bibitem[Crankshaw et~al\mbox{.}(2017)]%
        {crankshaw2017clipper}
\bibfield{author}{\bibinfo{person}{Daniel Crankshaw}, \bibinfo{person}{Xin Wang}, \bibinfo{person}{Guilio Zhou}, \bibinfo{person}{Michael~J Franklin}, \bibinfo{person}{Joseph~E Gonzalez}, {and} \bibinfo{person}{Ion Stoica}.} \bibinfo{year}{2017}\natexlab{}.
\newblock \showarticletitle{Clipper: A low-latency online prediction serving system}. In \bibinfo{booktitle}{\emph{14th $\{$USENIX$\}$ Symposium on Networked Systems Design and Implementation ($\{$NSDI$\}$ 17)}}. \bibinfo{pages}{613--627}.
\newblock


\bibitem[Datar et~al\mbox{.}(2004)]%
        {datar2004locality}
\bibfield{author}{\bibinfo{person}{Mayur Datar}, \bibinfo{person}{Nicole Immorlica}, \bibinfo{person}{Piotr Indyk}, {and} \bibinfo{person}{Vahab~S Mirrokni}.} \bibinfo{year}{2004}\natexlab{}.
\newblock \showarticletitle{Locality-sensitive hashing scheme based on p-stable distributions}. In \bibinfo{booktitle}{\emph{Proceedings of the twentieth annual symposium on Computational geometry}}. \bibinfo{pages}{253--262}.
\newblock


\bibitem[Desfontaines(2021)]%
        {desfontaines2021list}
\bibfield{author}{\bibinfo{person}{Damien Desfontaines}.} \bibinfo{year}{2021}\natexlab{}.
\newblock \showarticletitle{A list of real-world uses of differential privacy}.
\newblock \bibinfo{journal}{\emph{Ted is writing things}} (\bibinfo{year}{2021}).
\newblock


\bibitem[Doerr et~al\mbox{.}(2011)]%
        {doerr2011black}
\bibfield{author}{\bibinfo{person}{Benjamin Doerr}, \bibinfo{person}{Johannes Lengler}, \bibinfo{person}{Timo K{\"o}tzing}, {and} \bibinfo{person}{Carola Winzen}.} \bibinfo{year}{2011}\natexlab{}.
\newblock \showarticletitle{Black-box complexities of combinatorial problems}. In \bibinfo{booktitle}{\emph{Proceedings of the 13th annual conference on Genetic and evolutionary computation}}. \bibinfo{pages}{981--988}.
\newblock


\bibitem[Dosovitskiy et~al\mbox{.}(2021)]%
        {dosovitskiy2021an}
\bibfield{author}{\bibinfo{person}{Alexey Dosovitskiy}, \bibinfo{person}{Lucas Beyer}, \bibinfo{person}{Alexander Kolesnikov}, \bibinfo{person}{Dirk Weissenborn}, \bibinfo{person}{Xiaohua Zhai}, \bibinfo{person}{Thomas Unterthiner}, \bibinfo{person}{Mostafa Dehghani}, \bibinfo{person}{Matthias Minderer}, \bibinfo{person}{Georg Heigold}, \bibinfo{person}{Sylvain Gelly}, \bibinfo{person}{Jakob Uszkoreit}, {and} \bibinfo{person}{Neil Houlsby}.} \bibinfo{year}{2021}\natexlab{}.
\newblock \showarticletitle{An Image is Worth 16x16 Words: Transformers for Image Recognition at Scale}. In \bibinfo{booktitle}{\emph{International Conference on Learning Representations}}.
\newblock
\urldef\tempurl%
\url{https://openreview.net/forum?id=YicbFdNTTy}
\showURL{%
\tempurl}


\bibitem[Dwork(2008)]%
        {dwork2008differential}
\bibfield{author}{\bibinfo{person}{Cynthia Dwork}.} \bibinfo{year}{2008}\natexlab{}.
\newblock \showarticletitle{Differential privacy: A survey of results}. In \bibinfo{booktitle}{\emph{International conference on theory and applications of models of computation}}. Springer, \bibinfo{pages}{1--19}.
\newblock


\bibitem[Dwork and Roth(2014)]%
        {Dwork2014DP}
\bibfield{author}{\bibinfo{person}{Cynthia Dwork} {and} \bibinfo{person}{Aaron Roth}.} \bibinfo{year}{2014}\natexlab{}.
\newblock \showarticletitle{The Algorithmic Foundations of Differential Privacy}.
\newblock \bibinfo{journal}{\emph{Found. Trends Theor. Comput. Sci.}} \bibinfo{volume}{9}, \bibinfo{number}{3–4} (\bibinfo{date}{aug} \bibinfo{year}{2014}), \bibinfo{pages}{211–407}.
\newblock
\showISSN{1551-305X}
\urldef\tempurl%
\url{https://doi.org/10.1561/0400000042}
\showDOI{\tempurl}


\bibitem[Elmagarmid et~al\mbox{.}(2006)]%
        {elmagarmid2006duplicate}
\bibfield{author}{\bibinfo{person}{Ahmed~K Elmagarmid}, \bibinfo{person}{Panagiotis~G Ipeirotis}, {and} \bibinfo{person}{Vassilios~S Verykios}.} \bibinfo{year}{2006}\natexlab{}.
\newblock \showarticletitle{Duplicate record detection: A survey}.
\newblock \bibinfo{journal}{\emph{IEEE Transactions on knowledge and data engineering}} \bibinfo{volume}{19}, \bibinfo{number}{1} (\bibinfo{year}{2006}), \bibinfo{pages}{1--16}.
\newblock


\bibitem[Frantar and Alistarh(2022)]%
        {frantar2022optimal}
\bibfield{author}{\bibinfo{person}{Elias Frantar} {and} \bibinfo{person}{Dan Alistarh}.} \bibinfo{year}{2022}\natexlab{}.
\newblock \showarticletitle{Optimal brain compression: A framework for accurate post-training quantization and pruning}.
\newblock \bibinfo{journal}{\emph{Advances in Neural Information Processing Systems}}  \bibinfo{volume}{35} (\bibinfo{year}{2022}), \bibinfo{pages}{4475--4488}.
\newblock


\bibitem[Frantar et~al\mbox{.}(2022)]%
        {frantar2022gptq}
\bibfield{author}{\bibinfo{person}{Elias Frantar}, \bibinfo{person}{Saleh Ashkboos}, \bibinfo{person}{Torsten Hoefler}, {and} \bibinfo{person}{Dan Alistarh}.} \bibinfo{year}{2022}\natexlab{}.
\newblock \showarticletitle{Gptq: Accurate post-training quantization for generative pre-trained transformers}.
\newblock \bibinfo{journal}{\emph{arXiv preprint arXiv:2210.17323}} (\bibinfo{year}{2022}).
\newblock


\bibitem[Fredrikson et~al\mbox{.}(2015)]%
        {fredrikson2015model}
\bibfield{author}{\bibinfo{person}{Matt Fredrikson}, \bibinfo{person}{Somesh Jha}, {and} \bibinfo{person}{Thomas Ristenpart}.} \bibinfo{year}{2015}\natexlab{}.
\newblock \showarticletitle{Model inversion attacks that exploit confidence information and basic countermeasures}. In \bibinfo{booktitle}{\emph{Proceedings of the 22nd ACM SIGSAC conference on computer and communications security}}. \bibinfo{pages}{1322--1333}.
\newblock


\bibitem[Fu et~al\mbox{.}(2023)]%
        {fu2023dpsur}
\bibfield{author}{\bibinfo{person}{Jie Fu}, \bibinfo{person}{Qingqing Ye}, \bibinfo{person}{Haibo Hu}, \bibinfo{person}{Zhili Chen}, \bibinfo{person}{Lulu Wang}, \bibinfo{person}{Kuncan Wang}, {and} \bibinfo{person}{Ran Xun}.} \bibinfo{year}{2023}\natexlab{}.
\newblock \showarticletitle{Dpsur: Accelerating differentially private stochastic gradient descent using selective update and release}.
\newblock \bibinfo{journal}{\emph{arXiv preprint arXiv:2311.14056}} (\bibinfo{year}{2023}).
\newblock


\bibitem[Ghazi et~al\mbox{.}(2021)]%
        {ghazi2021deep}
\bibfield{author}{\bibinfo{person}{Badih Ghazi}, \bibinfo{person}{Noah Golowich}, \bibinfo{person}{Ravi Kumar}, \bibinfo{person}{Pasin Manurangsi}, {and} \bibinfo{person}{Chiyuan Zhang}.} \bibinfo{year}{2021}\natexlab{}.
\newblock \showarticletitle{Deep learning with label differential privacy}.
\newblock \bibinfo{journal}{\emph{Advances in neural information processing systems}}  \bibinfo{volume}{34} (\bibinfo{year}{2021}), \bibinfo{pages}{27131--27145}.
\newblock


\bibitem[Gholami et~al\mbox{.}(2022)]%
        {gholami2022survey}
\bibfield{author}{\bibinfo{person}{Amir Gholami}, \bibinfo{person}{Sehoon Kim}, \bibinfo{person}{Zhen Dong}, \bibinfo{person}{Zhewei Yao}, \bibinfo{person}{Michael~W Mahoney}, {and} \bibinfo{person}{Kurt Keutzer}.} \bibinfo{year}{2022}\natexlab{}.
\newblock \showarticletitle{A survey of quantization methods for efficient neural network inference}.
\newblock In \bibinfo{booktitle}{\emph{Low-Power Computer Vision}}. \bibinfo{publisher}{Chapman and Hall/CRC}, \bibinfo{pages}{291--326}.
\newblock


\bibitem[Gou et~al\mbox{.}(2021)]%
        {gou2021knowledge}
\bibfield{author}{\bibinfo{person}{Jianping Gou}, \bibinfo{person}{Baosheng Yu}, \bibinfo{person}{Stephen~J Maybank}, {and} \bibinfo{person}{Dacheng Tao}.} \bibinfo{year}{2021}\natexlab{}.
\newblock \showarticletitle{Knowledge distillation: A survey}.
\newblock \bibinfo{journal}{\emph{International Journal of Computer Vision}} \bibinfo{volume}{129}, \bibinfo{number}{6} (\bibinfo{year}{2021}), \bibinfo{pages}{1789--1819}.
\newblock


\bibitem[Hard et~al\mbox{.}(2018)]%
        {hard2018federated}
\bibfield{author}{\bibinfo{person}{Andrew Hard}, \bibinfo{person}{Kanishka Rao}, \bibinfo{person}{Rajiv Mathews}, \bibinfo{person}{Swaroop Ramaswamy}, \bibinfo{person}{Fran{\c{c}}oise Beaufays}, \bibinfo{person}{Sean Augenstein}, \bibinfo{person}{Hubert Eichner}, \bibinfo{person}{Chlo{\'e} Kiddon}, {and} \bibinfo{person}{Daniel Ramage}.} \bibinfo{year}{2018}\natexlab{}.
\newblock \showarticletitle{Federated learning for mobile keyboard prediction}.
\newblock \bibinfo{journal}{\emph{arXiv preprint arXiv:1811.03604}} (\bibinfo{year}{2018}).
\newblock


\bibitem[He et~al\mbox{.}(2016)]%
        {he2016deep}
\bibfield{author}{\bibinfo{person}{Kaiming He}, \bibinfo{person}{Xiangyu Zhang}, \bibinfo{person}{Shaoqing Ren}, {and} \bibinfo{person}{Jian Sun}.} \bibinfo{year}{2016}\natexlab{}.
\newblock \showarticletitle{Deep residual learning for image recognition}. In \bibinfo{booktitle}{\emph{Proceedings of the IEEE conference on computer vision and pattern recognition}}. \bibinfo{pages}{770--778}.
\newblock


\bibitem[Hern{\'a}ndez and Stolfo(1995)]%
        {hernandez1995merge}
\bibfield{author}{\bibinfo{person}{Mauricio~A Hern{\'a}ndez} {and} \bibinfo{person}{Salvatore~J Stolfo}.} \bibinfo{year}{1995}\natexlab{}.
\newblock \showarticletitle{The merge/purge problem for large databases}.
\newblock \bibinfo{journal}{\emph{ACM Sigmod Record}} \bibinfo{volume}{24}, \bibinfo{number}{2} (\bibinfo{year}{1995}), \bibinfo{pages}{127--138}.
\newblock


\bibitem[Hu et~al\mbox{.}(2021)]%
        {hu2021lora}
\bibfield{author}{\bibinfo{person}{Edward~J Hu}, \bibinfo{person}{Yelong Shen}, \bibinfo{person}{Phillip Wallis}, \bibinfo{person}{Zeyuan Allen-Zhu}, \bibinfo{person}{Yuanzhi Li}, \bibinfo{person}{Shean Wang}, \bibinfo{person}{Lu Wang}, {and} \bibinfo{person}{Weizhu Chen}.} \bibinfo{year}{2021}\natexlab{}.
\newblock \showarticletitle{Lora: Low-rank adaptation of large language models}.
\newblock \bibinfo{journal}{\emph{arXiv preprint arXiv:2106.09685}} (\bibinfo{year}{2021}).
\newblock


\bibitem[Hu et~al\mbox{.}(2022)]%
        {hu2022membership}
\bibfield{author}{\bibinfo{person}{Hongsheng Hu}, \bibinfo{person}{Zoran Salcic}, \bibinfo{person}{Lichao Sun}, \bibinfo{person}{Gillian Dobbie}, \bibinfo{person}{Philip~S Yu}, {and} \bibinfo{person}{Xuyun Zhang}.} \bibinfo{year}{2022}\natexlab{}.
\newblock \showarticletitle{Membership inference attacks on machine learning: A survey}.
\newblock \bibinfo{journal}{\emph{ACM Computing Surveys (CSUR)}} \bibinfo{volume}{54}, \bibinfo{number}{11s} (\bibinfo{year}{2022}), \bibinfo{pages}{1--37}.
\newblock


\bibitem[Huang et~al\mbox{.}(2023)]%
        {huang2023elastictrainer}
\bibfield{author}{\bibinfo{person}{Kai Huang}, \bibinfo{person}{Boyuan Yang}, {and} \bibinfo{person}{Wei Gao}.} \bibinfo{year}{2023}\natexlab{}.
\newblock \showarticletitle{Elastictrainer: Speeding up on-device training with runtime elastic tensor selection}. In \bibinfo{booktitle}{\emph{Proceedings of the 21st Annual International Conference on Mobile Systems, Applications and Services}}. \bibinfo{pages}{56--69}.
\newblock


\bibitem[Hunt et~al\mbox{.}(2018)]%
        {hunt2018chiron}
\bibfield{author}{\bibinfo{person}{Tyler Hunt}, \bibinfo{person}{Congzheng Song}, \bibinfo{person}{Reza Shokri}, \bibinfo{person}{Vitaly Shmatikov}, {and} \bibinfo{person}{Emmett Witchel}.} \bibinfo{year}{2018}\natexlab{}.
\newblock \showarticletitle{Chiron: Privacy-preserving machine learning as a service}.
\newblock \bibinfo{journal}{\emph{arXiv preprint arXiv:1803.05961}} (\bibinfo{year}{2018}).
\newblock


\bibitem[Jagielski et~al\mbox{.}(2020)]%
        {jagielski2020auditing}
\bibfield{author}{\bibinfo{person}{Matthew Jagielski}, \bibinfo{person}{Jonathan Ullman}, {and} \bibinfo{person}{Alina Oprea}.} \bibinfo{year}{2020}\natexlab{}.
\newblock \showarticletitle{Auditing differentially private machine learning: How private is private SGD?}
\newblock \bibinfo{journal}{\emph{Advances in Neural Information Processing Systems}}  \bibinfo{volume}{33} (\bibinfo{year}{2020}), \bibinfo{pages}{22205--22216}.
\newblock


\bibitem[Kolb et~al\mbox{.}(2012a)]%
        {kolb2012dedoop}
\bibfield{author}{\bibinfo{person}{Lars Kolb}, \bibinfo{person}{Andreas Thor}, {and} \bibinfo{person}{Erhard Rahm}.} \bibinfo{year}{2012}\natexlab{a}.
\newblock \showarticletitle{Dedoop: Efficient deduplication with hadoop}.
\newblock \bibinfo{journal}{\emph{Proceedings of the VLDB Endowment}} \bibinfo{volume}{5}, \bibinfo{number}{12} (\bibinfo{year}{2012}), \bibinfo{pages}{1878--1881}.
\newblock


\bibitem[Kolb et~al\mbox{.}(2012b)]%
        {kolb2012load}
\bibfield{author}{\bibinfo{person}{Lars Kolb}, \bibinfo{person}{Andreas Thor}, {and} \bibinfo{person}{Erhard Rahm}.} \bibinfo{year}{2012}\natexlab{b}.
\newblock \showarticletitle{Load balancing for mapreduce-based entity resolution}. In \bibinfo{booktitle}{\emph{2012 IEEE 28th international conference on data engineering}}. IEEE, \bibinfo{pages}{618--629}.
\newblock


\bibitem[Krizhevsky(2009)]%
        {Krizhevsky09learningmultiple}
\bibfield{author}{\bibinfo{person}{Alex Krizhevsky}.} \bibinfo{year}{2009}\natexlab{}.
\newblock \bibinfo{booktitle}{\emph{Learning multiple layers of features from tiny images}}.
\newblock \bibinfo{type}{{T}echnical {R}eport}.
\newblock


\bibitem[Krizhevsky et~al\mbox{.}(2009)]%
        {krizhevsky2009learning}
\bibfield{author}{\bibinfo{person}{Alex Krizhevsky}, \bibinfo{person}{Geoffrey Hinton}, {et~al\mbox{.}}} \bibinfo{year}{2009}\natexlab{}.
\newblock \showarticletitle{Learning multiple layers of features from tiny images}.
\newblock  (\bibinfo{year}{2009}).
\newblock


\bibitem[Lee et~al\mbox{.}(2020)]%
        {Lee2020A}
\bibfield{author}{\bibinfo{person}{Namhoon Lee}, \bibinfo{person}{Thalaiyasingam Ajanthan}, \bibinfo{person}{Stephen Gould}, {and} \bibinfo{person}{Philip H.~S. Torr}.} \bibinfo{year}{2020}\natexlab{}.
\newblock \showarticletitle{A Signal Propagation Perspective for Pruning Neural Networks at Initialization}. In \bibinfo{booktitle}{\emph{International Conference on Learning Representations}}.
\newblock
\urldef\tempurl%
\url{https://openreview.net/forum?id=HJeTo2VFwH}
\showURL{%
\tempurl}


\bibitem[Lee and Nirjon(2020)]%
        {lee2020fast}
\bibfield{author}{\bibinfo{person}{Seulki Lee} {and} \bibinfo{person}{Shahriar Nirjon}.} \bibinfo{year}{2020}\natexlab{}.
\newblock \showarticletitle{Fast and scalable in-memory deep multitask learning via neural weight virtualization}. In \bibinfo{booktitle}{\emph{Proceedings of the 18th International Conference on Mobile Systems, Applications, and Services}}. \bibinfo{pages}{175--190}.
\newblock


\bibitem[Li et~al\mbox{.}(2017)]%
        {li2017pruning}
\bibfield{author}{\bibinfo{person}{Hao Li}, \bibinfo{person}{Asim Kadav}, \bibinfo{person}{Igor Durdanovic}, \bibinfo{person}{Hanan Samet}, {and} \bibinfo{person}{Hans~Peter Graf}.} \bibinfo{year}{2017}\natexlab{}.
\newblock \showarticletitle{Pruning Filters for Efficient ConvNets}. In \bibinfo{booktitle}{\emph{International Conference on Learning Representations}}.
\newblock
\urldef\tempurl%
\url{https://openreview.net/forum?id=rJqFGTslg}
\showURL{%
\tempurl}


\bibitem[Lin and Kifer(2014)]%
        {lin2014arbitrage}
\bibfield{author}{\bibinfo{person}{Bing-Rong Lin} {and} \bibinfo{person}{Daniel Kifer}.} \bibinfo{year}{2014}\natexlab{}.
\newblock \showarticletitle{On arbitrage-free pricing for general data queries}.
\newblock \bibinfo{journal}{\emph{Proceedings of the VLDB Endowment}} \bibinfo{volume}{7}, \bibinfo{number}{9} (\bibinfo{year}{2014}), \bibinfo{pages}{757--768}.
\newblock


\bibitem[Lin et~al\mbox{.}(2024)]%
        {lin2023awq}
\bibfield{author}{\bibinfo{person}{Ji Lin}, \bibinfo{person}{Jiaming Tang}, \bibinfo{person}{Haotian Tang}, \bibinfo{person}{Shang Yang}, \bibinfo{person}{Wei-Ming Chen}, \bibinfo{person}{Wei-Chen Wang}, \bibinfo{person}{Guangxuan Xiao}, \bibinfo{person}{Xingyu Dang}, \bibinfo{person}{Chuang Gan}, {and} \bibinfo{person}{Song Han}.} \bibinfo{year}{2024}\natexlab{}.
\newblock \showarticletitle{AWQ: Activation-aware Weight Quantization for LLM Compression and Acceleration}. In \bibinfo{booktitle}{\emph{MLSys}}.
\newblock


\bibitem[Lin et~al\mbox{.}(2023)]%
        {lin2023smartlite}
\bibfield{author}{\bibinfo{person}{Qiuru Lin}, \bibinfo{person}{Sai Wu}, \bibinfo{person}{Junbo Zhao}, \bibinfo{person}{Jian Dai}, \bibinfo{person}{Meng Shi}, \bibinfo{person}{Gang Chen}, {and} \bibinfo{person}{Feifei Li}.} \bibinfo{year}{2023}\natexlab{}.
\newblock \showarticletitle{SmartLite: A DBMS-Based Serving System for DNN Inference in Resource-Constrained Environments}.
\newblock \bibinfo{journal}{\emph{Proceedings of the VLDB Endowment}} \bibinfo{volume}{17}, \bibinfo{number}{3} (\bibinfo{year}{2023}), \bibinfo{pages}{278--291}.
\newblock


\bibitem[Liu et~al\mbox{.}(2021a)]%
        {liu2021dealer}
\bibfield{author}{\bibinfo{person}{Jinfei Liu}, \bibinfo{person}{Jian Lou}, \bibinfo{person}{Junxu Liu}, \bibinfo{person}{Li Xiong}, \bibinfo{person}{Jian Pei}, {and} \bibinfo{person}{Jimeng Sun}.} \bibinfo{year}{2021}\natexlab{a}.
\newblock \showarticletitle{Dealer: an end-to-end model marketplace with differential privacy}.
\newblock \bibinfo{journal}{\emph{Proceedings of the VLDB Endowment}} \bibinfo{volume}{14}, \bibinfo{number}{6} (\bibinfo{year}{2021}).
\newblock


\bibitem[Liu et~al\mbox{.}(2021b)]%
        {liu2021group}
\bibfield{author}{\bibinfo{person}{Liyang Liu}, \bibinfo{person}{Shilong Zhang}, \bibinfo{person}{Zhanghui Kuang}, \bibinfo{person}{Aojun Zhou}, \bibinfo{person}{Jing-Hao Xue}, \bibinfo{person}{Xinjiang Wang}, \bibinfo{person}{Yimin Chen}, \bibinfo{person}{Wenming Yang}, \bibinfo{person}{Qingmin Liao}, {and} \bibinfo{person}{Wayne Zhang}.} \bibinfo{year}{2021}\natexlab{b}.
\newblock \showarticletitle{Group fisher pruning for practical network compression}. In \bibinfo{booktitle}{\emph{International Conference on Machine Learning}}. PMLR, \bibinfo{pages}{7021--7032}.
\newblock


\bibitem[Liu et~al\mbox{.}(2019)]%
        {liu2019roberta}
\bibfield{author}{\bibinfo{person}{Yinhan Liu}, \bibinfo{person}{Myle Ott}, \bibinfo{person}{Naman Goyal}, \bibinfo{person}{Jingfei Du}, \bibinfo{person}{Mandar Joshi}, \bibinfo{person}{Danqi Chen}, \bibinfo{person}{Omer Levy}, \bibinfo{person}{Mike Lewis}, \bibinfo{person}{Luke Zettlemoyer}, {and} \bibinfo{person}{Veselin Stoyanov}.} \bibinfo{year}{2019}\natexlab{}.
\newblock \showarticletitle{Roberta: A robustly optimized bert pretraining approach}.
\newblock \bibinfo{journal}{\emph{arXiv preprint arXiv:1907.11692}} (\bibinfo{year}{2019}).
\newblock


\bibitem[Liu et~al\mbox{.}(2015)]%
        {liu2015faceattributes}
\bibfield{author}{\bibinfo{person}{Ziwei Liu}, \bibinfo{person}{Ping Luo}, \bibinfo{person}{Xiaogang Wang}, {and} \bibinfo{person}{Xiaoou Tang}.} \bibinfo{year}{2015}\natexlab{}.
\newblock \showarticletitle{Deep Learning Face Attributes in the Wild}. In \bibinfo{booktitle}{\emph{Proceedings of International Conference on Computer Vision (ICCV)}}.
\newblock


\bibitem[Luo et~al\mbox{.}(2021)]%
        {luo2021privacy}
\bibfield{author}{\bibinfo{person}{Tao Luo}, \bibinfo{person}{Mingen Pan}, \bibinfo{person}{Pierre Tholoniat}, \bibinfo{person}{Asaf Cidon}, \bibinfo{person}{Roxana Geambasu}, {and} \bibinfo{person}{Mathias L{\'e}cuyer}.} \bibinfo{year}{2021}\natexlab{}.
\newblock \showarticletitle{Privacy budget scheduling}. In \bibinfo{booktitle}{\emph{15th $\{$USENIX$\}$ Symposium on Operating Systems Design and Implementation ($\{$OSDI$\}$ 21)}}. \bibinfo{pages}{55--74}.
\newblock


\bibitem[Lyu et~al\mbox{.}(2017)]%
        {lyu2017understanding}
\bibfield{author}{\bibinfo{person}{Min Lyu}, \bibinfo{person}{Dong Su}, {and} \bibinfo{person}{Ninghui Li}.} \bibinfo{year}{2017}\natexlab{}.
\newblock \showarticletitle{Understanding the Sparse Vector Technique for Differential Privacy}.
\newblock \bibinfo{journal}{\emph{Proceedings of the VLDB Endowment}} \bibinfo{volume}{10}, \bibinfo{number}{6} (\bibinfo{year}{2017}).
\newblock


\bibitem[Maas et~al\mbox{.}(2011)]%
        {maas2011learning}
\bibfield{author}{\bibinfo{person}{Andrew Maas}, \bibinfo{person}{Raymond~E Daly}, \bibinfo{person}{Peter~T Pham}, \bibinfo{person}{Dan Huang}, \bibinfo{person}{Andrew~Y Ng}, {and} \bibinfo{person}{Christopher Potts}.} \bibinfo{year}{2011}\natexlab{}.
\newblock \showarticletitle{Learning word vectors for sentiment analysis}. In \bibinfo{booktitle}{\emph{Proceedings of the 49th annual meeting of the association for computational linguistics: Human language technologies}}. \bibinfo{pages}{142--150}.
\newblock


\bibitem[Mao et~al\mbox{.}(2017)]%
        {mao2017s2jsd}
\bibfield{author}{\bibinfo{person}{Xian-Ling Mao}, \bibinfo{person}{Bo-Si Feng}, \bibinfo{person}{Yi-Jing Hao}, \bibinfo{person}{Liqiang Nie}, \bibinfo{person}{Heyan Huang}, {and} \bibinfo{person}{Guihua Wen}.} \bibinfo{year}{2017}\natexlab{}.
\newblock \showarticletitle{S2JSD-LSH: A locality-sensitive hashing schema for probability distributions}. In \bibinfo{booktitle}{\emph{Proceedings of the AAAI Conference on Artificial Intelligence}}, Vol.~\bibinfo{volume}{31}.
\newblock


\bibitem[McSherry and Mironov(2009)]%
        {mcsherry2009differentially}
\bibfield{author}{\bibinfo{person}{Frank McSherry} {and} \bibinfo{person}{Ilya Mironov}.} \bibinfo{year}{2009}\natexlab{}.
\newblock \showarticletitle{Differentially private recommender systems: Building privacy into the netflix prize contenders}. In \bibinfo{booktitle}{\emph{Proceedings of the 15th ACM SIGKDD international conference on Knowledge discovery and data mining}}. \bibinfo{pages}{627--636}.
\newblock


\bibitem[McSherry(2009a)]%
        {mcsherry2009privacy}
\bibfield{author}{\bibinfo{person}{Frank~D McSherry}.} \bibinfo{year}{2009}\natexlab{a}.
\newblock \showarticletitle{Privacy integrated queries: an extensible platform for privacy-preserving data analysis}. In \bibinfo{booktitle}{\emph{Proceedings of the 2009 ACM SIGMOD International Conference on Management of data}}. \bibinfo{pages}{19--30}.
\newblock


\bibitem[McSherry(2009b)]%
        {McSherry2009pquery}
\bibfield{author}{\bibinfo{person}{Frank~D. McSherry}.} \bibinfo{year}{2009}\natexlab{b}.
\newblock \showarticletitle{Privacy Integrated Queries: An Extensible Platform for Privacy-Preserving Data Analysis}. In \bibinfo{booktitle}{\emph{Proceedings of the 2009 ACM SIGMOD International Conference on Management of Data}} (Providence, Rhode Island, USA) \emph{(\bibinfo{series}{SIGMOD '09})}. \bibinfo{publisher}{Association for Computing Machinery}, \bibinfo{address}{New York, NY, USA}, \bibinfo{pages}{19–30}.
\newblock
\showISBNx{9781605585512}
\urldef\tempurl%
\url{https://doi.org/10.1145/1559845.1559850}
\showDOI{\tempurl}


\bibitem[MESTS and Tang(2018)]%
        {mests2018distributed}
\bibfield{author}{\bibinfo{person}{Joty MESTS} {and} \bibinfo{person}{MON Tang}.} \bibinfo{year}{2018}\natexlab{}.
\newblock \showarticletitle{Distributed representations of tuples for entity resolution}.
\newblock \bibinfo{journal}{\emph{Proceedings of the VLDB Endowment}} \bibinfo{volume}{11}, \bibinfo{number}{11} (\bibinfo{year}{2018}).
\newblock


\bibitem[Nanayakkara et~al\mbox{.}(2023)]%
        {nanayakkara2023chances}
\bibfield{author}{\bibinfo{person}{Priyanka Nanayakkara}, \bibinfo{person}{Mary~Anne Smart}, \bibinfo{person}{Rachel Cummings}, \bibinfo{person}{Gabriel Kaptchuk}, {and} \bibinfo{person}{Elissa~M Redmiles}.} \bibinfo{year}{2023}\natexlab{}.
\newblock \showarticletitle{What are the chances? explaining the epsilon parameter in differential privacy}. In \bibinfo{booktitle}{\emph{32nd USENIX Security Symposium (USENIX Security 23)}}. \bibinfo{pages}{1613--1630}.
\newblock


\bibitem[Netzer et~al\mbox{.}(2011)]%
        {netzer2011reading}
\bibfield{author}{\bibinfo{person}{Yuval Netzer}, \bibinfo{person}{Tao Wang}, \bibinfo{person}{Adam Coates}, \bibinfo{person}{Alessandro Bissacco}, \bibinfo{person}{Bo Wu}, {and} \bibinfo{person}{Andrew~Y Ng}.} \bibinfo{year}{2011}\natexlab{}.
\newblock \showarticletitle{Reading digits in natural images with unsupervised feature learning}.
\newblock  (\bibinfo{year}{2011}).
\newblock


\bibitem[Ramaswamy et~al\mbox{.}(2019)]%
        {ramaswamy2019learning}
\bibfield{author}{\bibinfo{person}{S Ramaswamy}, \bibinfo{person}{R Mathews}, \bibinfo{person}{K Rao}, {and} \bibinfo{person}{F Beaufays}.} \bibinfo{year}{2019}\natexlab{}.
\newblock \showarticletitle{Learning for Emoji Prediction in a Mobile Keyboard}.
\newblock \bibinfo{journal}{\emph{arXiv preprint arXiv:1906.04329}} (\bibinfo{year}{2019}).
\newblock


\bibitem[Rust and S{\o}gaard(2023)]%
        {rust2023differential}
\bibfield{author}{\bibinfo{person}{Phillip Rust} {and} \bibinfo{person}{Anders S{\o}gaard}.} \bibinfo{year}{2023}\natexlab{}.
\newblock \showarticletitle{Differential privacy, linguistic fairness, and training data influence: Impossibility and possibility theorems for multilingual language models}. In \bibinfo{booktitle}{\emph{International Conference on Machine Learning}}. PMLR, \bibinfo{pages}{29354--29387}.
\newblock


\bibitem[Shen et~al\mbox{.}(2019)]%
        {shen2019nexus}
\bibfield{author}{\bibinfo{person}{Haichen Shen}, \bibinfo{person}{Lequn Chen}, \bibinfo{person}{Yuchen Jin}, \bibinfo{person}{Liangyu Zhao}, \bibinfo{person}{Bingyu Kong}, \bibinfo{person}{Matthai Philipose}, \bibinfo{person}{Arvind Krishnamurthy}, {and} \bibinfo{person}{Ravi Sundaram}.} \bibinfo{year}{2019}\natexlab{}.
\newblock \showarticletitle{Nexus: A GPU cluster engine for accelerating DNN-based video analysis}. In \bibinfo{booktitle}{\emph{Proceedings of the 27th ACM Symposium on Operating Systems Principles}}. \bibinfo{pages}{322--337}.
\newblock


\bibitem[Shokri et~al\mbox{.}(2017)]%
        {shokri2017membership}
\bibfield{author}{\bibinfo{person}{Reza Shokri}, \bibinfo{person}{Marco Stronati}, \bibinfo{person}{Congzheng Song}, {and} \bibinfo{person}{Vitaly Shmatikov}.} \bibinfo{year}{2017}\natexlab{}.
\newblock \showarticletitle{Membership inference attacks against machine learning models}. In \bibinfo{booktitle}{\emph{2017 IEEE symposium on security and privacy (SP)}}. IEEE, \bibinfo{pages}{3--18}.
\newblock


\bibitem[Simonyan et~al\mbox{.}(2013)]%
        {Simonyan2013DeepIC}
\bibfield{author}{\bibinfo{person}{Karen Simonyan}, \bibinfo{person}{Andrea Vedaldi}, {and} \bibinfo{person}{Andrew Zisserman}.} \bibinfo{year}{2013}\natexlab{}.
\newblock \showarticletitle{Deep Inside Convolutional Networks: Visualising Image Classification Models and Saliency Maps}.
\newblock \bibinfo{journal}{\emph{CoRR}}  \bibinfo{volume}{abs/1312.6034} (\bibinfo{year}{2013}).
\newblock
\urldef\tempurl%
\url{https://api.semanticscholar.org/CorpusID:1450294}
\showURL{%
\tempurl}


\bibitem[snowflake({[n.\,d.]})]%
        {snowflake-privacy}
\bibfield{author}{\bibinfo{person}{snowflake}.} \bibinfo{year}{[n.\,d.]}\natexlab{}.
\newblock \bibinfo{title}{Differential Privacy in Snowflake}.
\newblock \bibinfo{howpublished}{\url{https://docs.snowflake.com/en/user-guide/diff-privacy/differential-privacy-overview\#label-diff-privacy-loss-budgets}}.
\newblock


\bibitem[Socher et~al\mbox{.}(2013)]%
        {socher2013recursive}
\bibfield{author}{\bibinfo{person}{Richard Socher}, \bibinfo{person}{Alex Perelygin}, \bibinfo{person}{Jean Wu}, \bibinfo{person}{Jason Chuang}, \bibinfo{person}{Christopher~D Manning}, \bibinfo{person}{Andrew~Y Ng}, {and} \bibinfo{person}{Christopher Potts}.} \bibinfo{year}{2013}\natexlab{}.
\newblock \showarticletitle{Recursive deep models for semantic compositionality over a sentiment treebank}. In \bibinfo{booktitle}{\emph{Proceedings of the 2013 conference on empirical methods in natural language processing}}. \bibinfo{pages}{1631--1642}.
\newblock


\bibitem[Stallkamp et~al\mbox{.}(2012)]%
        {stallkampManVsComputer2012}
\bibfield{author}{\bibinfo{person}{J. Stallkamp}, \bibinfo{person}{M. Schlipsing}, \bibinfo{person}{J. Salmen}, {and} \bibinfo{person}{C. Igel}.} \bibinfo{year}{2012}\natexlab{}.
\newblock \showarticletitle{Man vs. Computer: {{Benchmarking}} Machine Learning Algorithms for Traffic Sign Recognition}.
\newblock \bibinfo{journal}{\emph{Neural Networks}}  \bibinfo{volume}{32} (\bibinfo{date}{Aug.} \bibinfo{year}{2012}), \bibinfo{pages}{323--332}.
\newblock
\showISSN{0893-6080}
\urldef\tempurl%
\url{https://doi.org/10.1016/j.neunet.2012.02.016}
\showDOI{\tempurl}


\bibitem[Sun et~al\mbox{.}(2024)]%
        {sun2024simple}
\bibfield{author}{\bibinfo{person}{Mingjie Sun}, \bibinfo{person}{Zhuang Liu}, \bibinfo{person}{Anna Bair}, {and} \bibinfo{person}{J~Zico Kolter}.} \bibinfo{year}{2024}\natexlab{}.
\newblock \showarticletitle{A Simple and Effective Pruning Approach for Large Language Models}. In \bibinfo{booktitle}{\emph{The Twelfth International Conference on Learning Representations}}.
\newblock
\urldef\tempurl%
\url{https://openreview.net/forum?id=PxoFut3dWW}
\showURL{%
\tempurl}


\bibitem[Thirumuruganathan et~al\mbox{.}(2021)]%
        {thirumuruganathan2021deep}
\bibfield{author}{\bibinfo{person}{Saravanan Thirumuruganathan}, \bibinfo{person}{Han Li}, \bibinfo{person}{Nan Tang}, \bibinfo{person}{Mourad Ouzzani}, \bibinfo{person}{Yash Govind}, \bibinfo{person}{Derek Paulsen}, \bibinfo{person}{Glenn Fung}, {and} \bibinfo{person}{AnHai Doan}.} \bibinfo{year}{2021}\natexlab{}.
\newblock \showarticletitle{Deep learning for blocking in entity matching: a design space exploration}.
\newblock \bibinfo{journal}{\emph{Proceedings of the VLDB Endowment}} \bibinfo{volume}{14}, \bibinfo{number}{11} (\bibinfo{year}{2021}), \bibinfo{pages}{2459--2472}.
\newblock


\bibitem[Tramer and Boneh(2021)]%
        {tramer2021differentially}
\bibfield{author}{\bibinfo{person}{Florian Tramer} {and} \bibinfo{person}{Dan Boneh}.} \bibinfo{year}{2021}\natexlab{}.
\newblock \showarticletitle{Differentially Private Learning Needs Better Features (or Much More Data)}. In \bibinfo{booktitle}{\emph{International Conference on Learning Representations}}.
\newblock
\urldef\tempurl%
\url{https://openreview.net/forum?id=YTWGvpFOQD-}
\showURL{%
\tempurl}


\bibitem[Vartak et~al\mbox{.}(2018)]%
        {vartak2018mistique}
\bibfield{author}{\bibinfo{person}{Manasi Vartak}, \bibinfo{person}{Joana~M F.~da Trindade}, \bibinfo{person}{Samuel Madden}, {and} \bibinfo{person}{Matei Zaharia}.} \bibinfo{year}{2018}\natexlab{}.
\newblock \showarticletitle{Mistique: A system to store and query model intermediates for model diagnosis}. In \bibinfo{booktitle}{\emph{Proceedings of the 2018 International Conference on Management of Data}}. \bibinfo{pages}{1285--1300}.
\newblock


\bibitem[Wang et~al\mbox{.}(2018)]%
        {wang2018glue}
\bibfield{author}{\bibinfo{person}{Alex Wang}, \bibinfo{person}{Amanpreet Singh}, \bibinfo{person}{Julian Michael}, \bibinfo{person}{Felix Hill}, \bibinfo{person}{Omer Levy}, {and} \bibinfo{person}{Samuel~R Bowman}.} \bibinfo{year}{2018}\natexlab{}.
\newblock \showarticletitle{GLUE: A multi-task benchmark and analysis platform for natural language understanding}.
\newblock \bibinfo{journal}{\emph{arXiv preprint arXiv:1804.07461}} (\bibinfo{year}{2018}).
\newblock


\bibitem[Williams et~al\mbox{.}(2018)]%
        {williams2018broad}
\bibfield{author}{\bibinfo{person}{Adina Williams}, \bibinfo{person}{Nikita Nangia}, {and} \bibinfo{person}{Samuel Bowman}.} \bibinfo{year}{2018}\natexlab{}.
\newblock \showarticletitle{A Broad-Coverage Challenge Corpus for Sentence Understanding through Inference}. In \bibinfo{booktitle}{\emph{Proceedings of the 2018 Conference of the North {A}merican Chapter of the Association for Computational Linguistics: Human Language Technologies, Volume 1 (Long Papers)}}. \bibinfo{publisher}{Association for Computational Linguistics}, \bibinfo{address}{New Orleans, Louisiana}, \bibinfo{pages}{1112--1122}.
\newblock
\urldef\tempurl%
\url{https://doi.org/10.18653/v1/N18-1101}
\showDOI{\tempurl}


\bibitem[Xiao et~al\mbox{.}(2008)]%
        {xiao2008ed}
\bibfield{author}{\bibinfo{person}{Chuan Xiao}, \bibinfo{person}{Wei Wang}, {and} \bibinfo{person}{Xuemin Lin}.} \bibinfo{year}{2008}\natexlab{}.
\newblock \showarticletitle{Ed-join: an efficient algorithm for similarity joins with edit distance constraints}.
\newblock \bibinfo{journal}{\emph{Proceedings of the VLDB Endowment}} \bibinfo{volume}{1}, \bibinfo{number}{1} (\bibinfo{year}{2008}), \bibinfo{pages}{933--944}.
\newblock


\bibitem[Xu and Zhang(2017)]%
        {xu2017advances}
\bibfield{author}{\bibinfo{person}{Zheng Xu} {and} \bibinfo{person}{Yanxiang Zhang}.} \bibinfo{year}{2017}\natexlab{}.
\newblock \bibinfo{title}{Advances in private training for production on-device language models}.
\newblock
\newblock


\bibitem[Yang et~al\mbox{.}(2024)]%
        {yang2024model}
\bibfield{author}{\bibinfo{person}{Enneng Yang}, \bibinfo{person}{Li Shen}, \bibinfo{person}{Guibing Guo}, \bibinfo{person}{Xingwei Wang}, \bibinfo{person}{Xiaochun Cao}, \bibinfo{person}{Jie Zhang}, {and} \bibinfo{person}{Dacheng Tao}.} \bibinfo{year}{2024}\natexlab{}.
\newblock \showarticletitle{Model merging in llms, mllms, and beyond: Methods, theories, applications and opportunities}.
\newblock \bibinfo{journal}{\emph{arXiv preprint arXiv:2408.07666}} (\bibinfo{year}{2024}).
\newblock


\bibitem[Yin et~al\mbox{.}(2024)]%
        {yin2024llm}
\bibfield{author}{\bibinfo{person}{Wangsong Yin}, \bibinfo{person}{Mengwei Xu}, \bibinfo{person}{Yuanchun Li}, {and} \bibinfo{person}{Xuanzhe Liu}.} \bibinfo{year}{2024}\natexlab{}.
\newblock \showarticletitle{Llm as a system service on mobile devices}.
\newblock \bibinfo{journal}{\emph{arXiv preprint arXiv:2403.11805}} (\bibinfo{year}{2024}).
\newblock


\bibitem[Yu et~al\mbox{.}(2016)]%
        {yu2016generic}
\bibfield{author}{\bibinfo{person}{Chenyun Yu}, \bibinfo{person}{Sarana Nutanong}, \bibinfo{person}{Hangyu Li}, \bibinfo{person}{Cong Wang}, {and} \bibinfo{person}{Xingliang Yuan}.} \bibinfo{year}{2016}\natexlab{}.
\newblock \showarticletitle{A generic method for accelerating LSH-based similarity join processing}.
\newblock \bibinfo{journal}{\emph{IEEE Transactions on Knowledge and Data Engineering}} \bibinfo{volume}{29}, \bibinfo{number}{4} (\bibinfo{year}{2016}), \bibinfo{pages}{712--726}.
\newblock


\bibitem[Yu et~al\mbox{.}(2021a)]%
        {yu2021not}
\bibfield{author}{\bibinfo{person}{Da Yu}, \bibinfo{person}{Huishuai Zhang}, \bibinfo{person}{Wei Chen}, {and} \bibinfo{person}{Tie-Yan Liu}.} \bibinfo{year}{2021}\natexlab{a}.
\newblock \showarticletitle{Do not let privacy overbill utility: Gradient embedding perturbation for private learning}.
\newblock \bibinfo{journal}{\emph{arXiv preprint arXiv:2102.12677}} (\bibinfo{year}{2021}).
\newblock


\bibitem[Yu et~al\mbox{.}(2021b)]%
        {yu2021large}
\bibfield{author}{\bibinfo{person}{Da Yu}, \bibinfo{person}{Huishuai Zhang}, \bibinfo{person}{Wei Chen}, \bibinfo{person}{Jian Yin}, {and} \bibinfo{person}{Tie-Yan Liu}.} \bibinfo{year}{2021}\natexlab{b}.
\newblock \showarticletitle{Large scale private learning via low-rank reparametrization}. In \bibinfo{booktitle}{\emph{International Conference on Machine Learning}}. PMLR, \bibinfo{pages}{12208--12218}.
\newblock


\bibitem[Zhang and He(2023)]%
        {zhang2023dprovdb}
\bibfield{author}{\bibinfo{person}{Shufan Zhang} {and} \bibinfo{person}{Xi He}.} \bibinfo{year}{2023}\natexlab{}.
\newblock \showarticletitle{DProvDB: Differentially private query processing with Multi-Analyst provenance}.
\newblock \bibinfo{journal}{\emph{Proceedings of the ACM on Management of Data}} \bibinfo{volume}{1}, \bibinfo{number}{4} (\bibinfo{year}{2023}), \bibinfo{pages}{1--27}.
\newblock


\bibitem[Zhou et~al\mbox{.}(2022)]%
        {zhou2022serving}
\bibfield{author}{\bibinfo{person}{Lixi Zhou}, \bibinfo{person}{Jiaqing Chen}, \bibinfo{person}{Amitabh Das}, \bibinfo{person}{Hong Min}, \bibinfo{person}{Lei Yu}, \bibinfo{person}{Ming Zhao}, {and} \bibinfo{person}{Jia Zou}.} \bibinfo{year}{2022}\natexlab{}.
\newblock \showarticletitle{Serving deep learning models with deduplication from relational databases}.
\newblock \bibinfo{journal}{\emph{Proc. VLDB Endow.}} \bibinfo{volume}{15}, \bibinfo{number}{10} (\bibinfo{date}{jun} \bibinfo{year}{2022}), \bibinfo{pages}{2230–2243}.
\newblock
\showISSN{2150-8097}
\urldef\tempurl%
\url{https://doi.org/10.14778/3547305.3547325}
\showDOI{\tempurl}


\bibitem[Zhou et~al\mbox{.}(2024)]%
        {zhou2024serving}
\bibfield{author}{\bibinfo{person}{Lixi Zhou}, \bibinfo{person}{Qi Lin}, \bibinfo{person}{Kanchan Chowdhury}, \bibinfo{person}{Saif Masood}, \bibinfo{person}{Alexandre Eichenberger}, \bibinfo{person}{Hong Min}, \bibinfo{person}{Alexander Sim}, \bibinfo{person}{Jie Wang}, \bibinfo{person}{Yida Wang}, \bibinfo{person}{Kesheng Wu}, {et~al\mbox{.}}} \bibinfo{year}{2024}\natexlab{}.
\newblock \showarticletitle{Serving Deep Learning Models from Relational Databases}.
\newblock \bibinfo{journal}{\emph{Advances in Database Technology-EDBT}} \bibinfo{volume}{27}, \bibinfo{number}{3} (\bibinfo{year}{2024}), \bibinfo{pages}{717--724}.
\newblock


\end{thebibliography}
